\documentclass[epj,nopacs]{svjour}
\usepackage{tabularx}
\usepackage{arydshln}
\usepackage[switch,displaymath]{lineno} 
\usepackage{latexsym}  \usepackage{graphicx}  \usepackage{esvect}
\usepackage{amsmath}  \usepackage{mathabx} 
\usepackage{hepparticles} 
\usepackage{hepnames} 
\usepackage[alsoload=hep]{siunitx}
\usepackage{mathtools}
\usepackage{multirow} 

\begin{document}
\title{Measurement of polarization observables of the associated strangeness production in proton proton interactions}
\titlerunning {Polarization observables of the reaction pp $\rightarrow$ pK$^{+}\mathrm{\Lambda}$}
\subtitle{The COSY-TOF Collaboration}
\author{ F.~Hauenstein\inst{1,4} \and
  E.~Borodina\inst{1}\and
  H.~Clement\inst{5,6} \and
  E.~Doroshkevich\inst{5,6}\thanks{current address: Institute for Nuclear Research Moscow 117312, Russia} \and
  R.~Dzhygadlo\inst{1}\thanks {current address: Hadron Physics I, GSI Helmholtzzentrum f\"{u}r Schwerionenforschung GmbH}\and
  K.~Ehrhardt\inst{5,6} \and
  W.~Eyrich\inst{4} \and
  W.~Gast\inst{1} \and
  A.~Gillitzer\inst{1} \and
  D.~Grzonka\inst{1} \and
  S.~Jowzaee\inst{1,7} \and
  P.~Klaja\inst{1,4} \and
  L.~Kober\inst{4} \and
  K.~Kilian\inst{1} \and
  M.~Krapp\inst{4} \and
  M.~Mertens\inst{1}\thanks {current address: Universit\"{a}t Duisburg-Essen 45141 Essen, Germany}\and
  P.~Moskal\inst{7}\and
  J.~Ritman\inst{1,2,8} \and
  E.~Roderburg\inst{1}\thanks{corresponding author e.roderburg@fz-juelich.de} \and
  M.~R\"{o}der\inst{1}\thanks {current address: Corporate Development, Forschungszentrum J\"{u}lich, 52428 J\"{u}lich, Germany} \and
  W.~Schroeder\inst{9} \and
  T.~Sefzick\inst{1} \and
  J.~Smyrski\inst{7} \and
  P.~Wintz\inst{1} \and
  P.~W\"{u}stner\inst{3}}

\institute{
  Institut f\"{u}r Kernphysik, Forschungszentrum J\"{u}lich, 52428 J\"{u}lich, Germany \and 
  J\"{u}lich Aachen Research Alliance, Forces and Matter Experiments  (JARA-FAME) \and 
  Zentralinstitut f\"{u}r Engineering, Elektronik und Analytik, 52428 J\"{u}lich, Germany \and 
  Friedrich-Alexander-Universit\"{a}t Erlangen-N\"{u}rnberg, 91058 Erlangen, Germany \and 
  Physikalisches Institut der Universit\"{a}t T\"{u}bingen, Auf der Morgenstelle 14, 72076 T\"{u}bingen, Germany  \and 
  Kepler Center for Astro and Particle Physics, University of T\"{u}bingen, Auf der Morgenstelle 14, 72076 T\"{u}bingen, Germany \and 
  Institute of Physics, Jagellonian University, PL-30-348 Cracow, Poland \and 
  Experimentalphysik I, Ruhr-Universit\"{a}t Bochum, 44780 Bochum, Germany \and 
  Corporate Development, Forschungszentrum J\"{u}lich, 52428 J\"{u}lich, Germany 
}
\date{\today}
\authorrunning{The COSY-TOF Collaboration}

\abstract
{The $\mathrm{\Lambda}$ polarization, the analyzing power, and the
  $\mathrm{\Lambda}$ spin transfer coefficient of the reaction pp
  $\rightarrow$ pK$^{+}\mathrm{\Lambda}$ were measured at beam momenta
  of 2.70\,GeV/c and 2.95\,GeV/c corresponding to excess energies of
  122\,MeV and 204\,MeV.  While the analyzing power and the spin
  transfer coefficient do not change significantly with the excess
  energy, the $\mathrm{\Lambda}$ polarization varies strongly and
  changes its sign. As this is the first measurement of polarization
  observables below an excess energy of 200\,MeV, the change of the
  sign of the $\mathrm{\Lambda}$ polarization was not observed
  before. The high statistics of the data ($\approx$ 200\,k events for
  each momentum) enables detailed studies of the dependence of the
  $\mathrm{\Lambda}$ polarization and the analyzing power on the
  center of mass momentum of the particles. The results of the spin
  transfer coefficient are in qualitative agreement with the DISTO
  experiment. The $\Lambda$ polarization data of 2.95 GeV/c are only
  conform with the DISTO experiment, while both the 2.70 GeV/c and
  2.95 GeV/c data differ strongly from all previous measurements,
  whether exclusive or inclusive.}
 
\maketitle
\section{Introduction}

In this paper the results concerning the polarization observables of
recent COSY-TOF measurements of pp\,$\rightarrow$ pK$^{+}$$\mathrm{\Lambda}$
are presented. Angular distributions, Dalitz plots, and invariant
masses  were discussed in a previous
paper \cite {Jowzaee2016}.   Due to the nearly 4$\pi$
acceptance of the COSY-TOF detector for this reaction it is possible
to measure  the $\mathrm{\Lambda}$
polarization, the analyzing power determined with the final state particles, and
the  $\mathrm{\Lambda}$ spin transfer coefficient for the whole kinematic range.
For measurements of the analyzing power and the spin transfer coefficient
the experiment made use of the polarized extracted proton beam from the COSY accelerator, with
polarization up to 90\%

Since the first observation that $\mathrm{\Lambda}$'s exhibit a
polarization, even being produced with an unpolarized beam \cite
{Bunce1976}, many experiments examined the dependence of this
polarization on different kinematic variables. These experiments were
performed with high beam momenta.  No consistent behavior of the
$\mathrm{\Lambda}$ polarization emerged from these
measurements. 
Closer to threshold only inclusive data from HADES (p$_{\mathrm{b}}$ = 4.34\,GeV/c) \cite {Agakishiev2014} and exclusive data from DISTO
(p$_{\mathrm{b}}$ = 3.67 GeV/c) \cite {Choi1998} exist. 
The COSY-TOF data discussed in
this paper are the first to give information of the
$\mathrm{\Lambda}$ polarization as close as $\epsilon$ = 122\,MeV
above threshold. They exhibit a strong dependence of the
$\mathrm{\Lambda}$ polarization on the excess energy, which was not
observed before.

The analyzing power measured with the final state particles provides information on
the angular momenta involved in the process in addition to the information
that can be gained from the angular distributions. The measured distributions
should help to determine parameters of a partial wave analysis, which is in 
preparation \cite {Muenzer2015}. As most of the previous measurements
were inclusive and high acceptance is needed for measurements of 
the analyzing power, the existing data are very scarce and partly still
preliminary. As the COSY-TOF data cover the full accessible phase space, they can be described
with Legendre polynomials. The dependence of the Legendre coefficients
on the particle momenta is examined.

The measurement of the spin transfer coefficient combines information
on the $\mathrm{\Lambda}$ polarization with the analyzing power by an
event to event analysis.  Apart from three measurements at high
momenta, which were restricted in the kinematic range \cite
{Bonner1987b,Bonner1988,Bravar1997}, the only exclusive measurements
were published by the DISTO experiment \cite {Balestra1999a}.  The
COSY-TOF data allow for the first time a comparison of this observable
between independent measurements.

\section{Experimental setup and analysis}
The data were taken with the COSY-TOF spectrometer, which was situated
at an external beam of the accelerator COSY/J\"{u}lich. 
\begin{figure}[thb]
	\begin {center}
	\includegraphics[width=.5\textwidth]{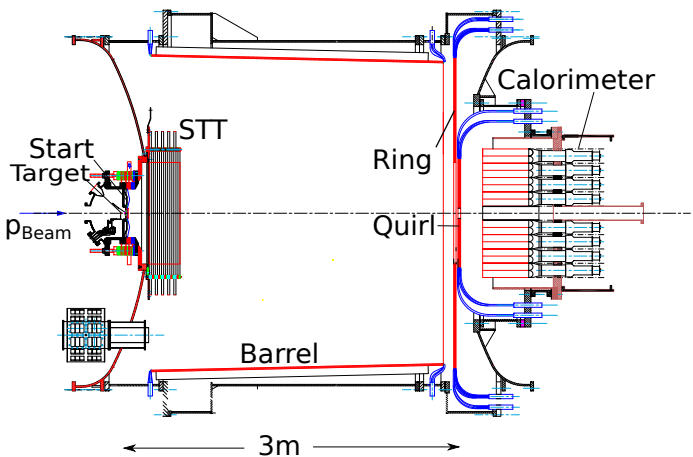}
	\caption{\label{experiment} Side view of the COSY-TOF
          spectrometer. In beam direction the start counter (Start),
          the straw tube tracker (STT) the barrel scintillators, the
          inner ring (Quirl), the outer ring (Ring) and the
          Calorimeter are shown.  All detectors and the liquid
          hydrogen target are located inside the vacuum vessel.
        }  
        \end {center}
\end{figure}
The COSY-TOF detector is a non magnetic spectrometer built as a
cylindrical vacuum vessel, with  3.5\,m length and 3\,m diameter (see
fig.\ref {experiment}). The inner walls are covered with segmented
scintillation counters, which are used for trigger signal generation.  The main components
are the miniaturized liquid hydrogen target in cylinder form with 6 mm
diameter and 4 mm length and the straw tube tracker, from which
precise information of the vertices and track directions are
obtained. The detector is described in detail in \cite {Jowzaee2016}.

The analysis of the pK$^+\mathrm{\Lambda}$ final state examines events 
which are triggered by four or more hits in the scintillator hodoscopes. These
events are fitted by a two vertices hypothesis: a primary vertex of
pK$^+$ and a secondary vertex of the $\mathrm{\Lambda}$ decay. Events for which this
fit converges are submitted to a kinematic fit procedure, which applies
momentum and energy conservation and the masses of the involved
particles. The assignment of the proton and kaon masses to the primary
tracks is done by applying the kinematic fit for each possibility, and the
result with the lowest $\chi^2$ is chosen \cite {Hauenstein2014,Jowzaee2014}.  
For the definition of a pK$\mathrm{\Lambda}$ event the following criteria 
are applied:  The $\chi^2$/NDF of the kinematic fit has to be less than 5, the decay length in the rest system
of the $\mathrm{\Lambda}$ has to be  larger than $\SI{2}{\centi\metre}$ and the angle of the decay proton
to the $\mathrm{\Lambda}$ direction has to be larger than \ang{2.5}.

The background introduced by multi-pion events is determined to be less  than 5\%
by comparing the measured $\mathrm{\Lambda}$ decay length distribution to the expected distribution
\cite{Jowzaee2016}. Special care has to be taken for the admixture of  
$\mathrm{\Lambda}$'s from pp\,$\rightarrow$ pK($\mathrm{\Sigma}^{0} \rightarrow \mathrm{\gamma}
\mathrm{\Lambda}$), which dilutes the measured $\mathrm{\Lambda}$ polarization and introduces
unknown errors to the spin transfer coefficient and to the analyzing
power. The suppression of the $\mathrm{\Sigma}^0$ events by the $\chi^2$ threshold for
the kinematic fit is studied with a Monte Carlo sample of
pK($\mathrm{\Sigma}^0\,\rightarrow$ $\mathrm{\gamma}\mathrm{\Lambda}$) events. Including the
cross section ratios, the contamination of $\mathrm{\Sigma}^0$ events is less
than 5\% for the 2.95\,GeV/c data \cite {Jowzaee2016} and less than 1\%  for the 2.70
GeV/c data \cite{Hauenstein2014}.  From the Monte Carlo
simulations it is known that the pp\,$\rightarrow$ pK($\mathrm{\Sigma}^0$
$\rightarrow$ $\mathrm{\gamma}$$\mathrm{\Lambda}$) events are shifted by the kinematic
fit, which assumes a pp\,$\rightarrow$ pK$\mathrm{\Lambda}$ reaction, to
backward kaon directions. Therefore, the 2.95\,GeV/c data are analyzed
in addition with a selection on the kaon angle of $|\mathrm{cos}\vartheta_{\mathrm{K}}^{\mathrm{cm}}| <
0.9$ and compared with the results without this restriction. Apart from
statistical fluctuations, no differences are found in the distributions
of all polarization observables.
 
In order to determine the beam polarization, a sample of elastic
scattering events is recorded by a trigger which requires at least two
charged tracks.  With cuts on the coplanarity and on the missing
energy elastic scattered events are determined with a background of
less than 1\% \cite {Hauenstein2014}.  By evaluating the left-right
asymmetry of the elastic scattered protons and by comparing this
distribution with the analyzing power determined with the partial wave
analysis SAID \cite {Arndt2007b} the averaged transverse beam
polarization is determined. For the 2.70\,GeV/c data it is (79.0 $\pm$
1.1)\% \cite {Hauenstein2014}. The 2.95\,GeV/c data were acquired in
two different runs, the first with 54,000 events had a polarization of
(61.0 $\pm$ 1.7)\% \cite {Roeder2013} and the second with 121,000
events had a polarization of (87.5 $\pm$ 2.0)\% \cite
{Jowzaee2014}. Both runs are analyzed together by assuming a weighted
mean beam polarization of (79.3 $\pm$ 2.0)\%.
For the data taking the direction of the beam polarization was changed
by switching the polarization in the H$^{-}$ ion source \cite {Gebel2008}
with every spill, which has a typical length of 100\,s.

The systematic errors due to instrumental asymmetries are studied by comparing the
results of the observables obtained by different methods: in case of
the $\mathrm{\Lambda}$ polarization these are the integral method, 
which is applied in this analysis (see  eq.
\ref{eq:hyperonpolarization2}), and 
the weighted sum method, which is described in \cite{Besset1979b}. 
In case of the analyzing power these are 
the double difference method (eq. 9 of reference \cite {Ohlsen1973}),
and the method described in eq. \ref{eq:analyzingpower}.
The comparison of both methods indicates that the deviations are within  
the range of their statistical error.
 Therefore, we assume that the systematical errors are less than 
or equal to  the statistical errors. For the spin transfer
measurement we assume the same systematical error due to instrumental asymmetries
 as in the former observables. The effect of
admixtures of pK($\mathrm{\Sigma}^0$\,$\rightarrow$ $\mathrm{\gamma}
\mathrm{\Lambda}$) events is expected to introduce the same amount of
systematic errors. With these assumptions the mean absolute systematic error
for the $\mathrm{\Lambda}$ polarization is 0.04.
For the measurement of the analyzing power and the spin transfer
coefficient the uncertainty of the beam polarization has to be added
to the systematic errors, the mean value of these systematic errors
is 0.05 (absolute value). 

\section{$\mathrm{\Lambda}$ polarization $P_{\mathrm{N}}$}
\subsection{Results}
The $\mathrm{\Lambda}$ polarization is given by the equation:
\begin{equation}\label{eq:hyperonpolarization1}
I(\theta^*) =  I_0\cdot(1+P_{\mathrm{N}}\: \alpha \cos(\theta^*))
\end{equation}
 $\theta^*$ is the angle between the direction of the decay proton (in the $\mathrm{\Lambda}$
rest frame) and the normal vector to the plane which is spanned by the beam proton and the $\mathrm{\Lambda}$ direction.
 $\alpha$ is the hyperon decay asymmetry parameter:
\newline
$\alpha(\mathrm{\Lambda}\rightarrow p \pi^{-}) = 0.642 \pm 0.013 $ \cite {Olive2014}

From eq. \ref{eq:hyperonpolarization1} the  $\mathrm{\Lambda}$ polarization is calculated by applying the 
difference of the count rates with  cos$(\theta^*)\,>\,0$  ($N_\mathrm{A}$) and   cos$(\theta^*)\,<\,0$ ($N_\mathrm{B}$):
\begin{equation}\label{eq:hyperonpolarization2}
P_{\mathrm{N}} = \frac{2} {\alpha} \frac{N_{\mathrm{A}}-N_{\mathrm{B}}}{N_{\mathrm{A}}+N_{\mathrm{B}}}
\end{equation}

\begin{figure}[htpb!]
	\begin {center}
	\includegraphics[width=.5\textwidth]{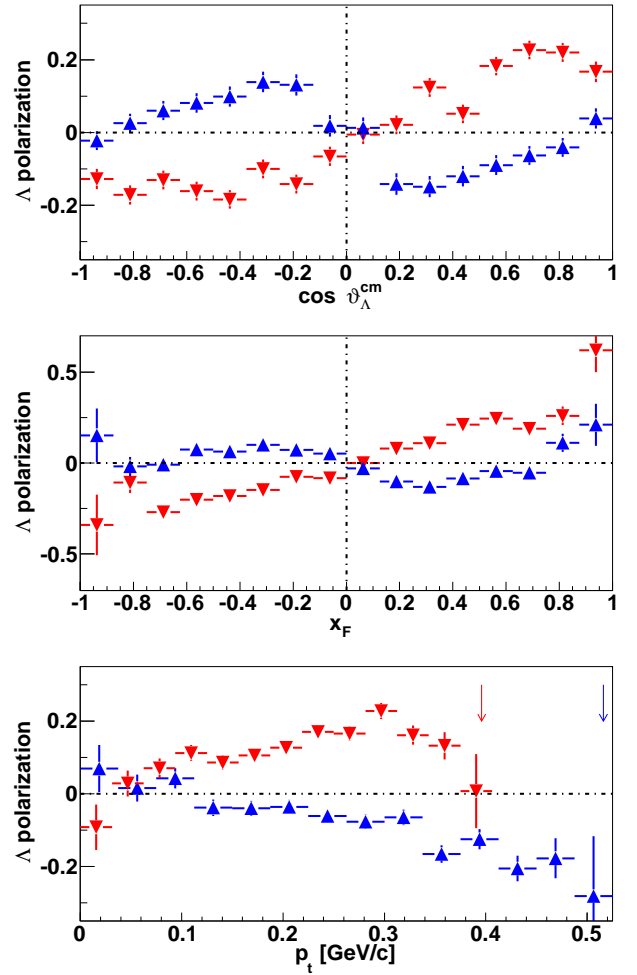}
	\caption{\label{lambdapolarization} 
          The $\mathrm{\Lambda}$ polarization is shown for two beam momenta:
          2.70\,GeV/c (red triangles down) and 2.95\,GeV/c (blue triangles
          up). Top: dependence on the $\mathrm{\Lambda}$ cm scattering
          angle. Middle: dependence on Feynman~$x_{\mathrm{F}}$. Bottom:
          dependence on the $\mathrm{\Lambda}$ transverse momentum. For each
          beam momentum  the maximum of the transverse momentum is
          indicated with arrows in the upper part of the bottom panel.
        }  
        \end {center}
\end{figure}

For the beam momentum of 2.70\,GeV/c 227,000 events and for the beam
momentum of 2.95\,GeV/c 206,000 events are analyzed, including runs without
beam polarization.  The $\mathrm{\Lambda}$ polarization
is shown in fig. \ref {lambdapolarization} as a function of the scattering
angle, of Feynman~$x_{\mathrm{F}}$\footnote{Feynman~$x_{\mathrm{F}}$
  is defined as the ratio of the longitudinal cm momentum to its
  maximum possible value}, and of the transverse momentum.
The initial system consists of two identical
particles\footnote{     this is exactly true only for an unpolarized beam,
                        the polarization of the beam proton violates this symmetry. But 
                        the $\mathrm{\Lambda}$ polarization is evaluated by taking the average of the
                        up and down polarization runs. The integrated luminosities
                        with up and down polarization are almost the same.},                        
therefore, the $\mathrm{\Lambda}$ polarization has to
change its sign between the backward $\mathrm{\Lambda}$ region and the forward
$\mathrm{\Lambda}$ region:

\begin{eqnarray}\label{forwardbackward}
P_{ \mathrm{\Lambda}}(\mathrm{cos}(\vartheta_{\mathrm{\Lambda}}^{\mathrm{cm}})) & = & - P_{\mathrm{\Lambda}}(-\mathrm{cos}(\vartheta_{\mathrm{\Lambda}}^{\mathrm{cm}})) \\
P_{ \mathrm{\Lambda}}(x_{\mathrm{F}}) &                    = & - P_{ \mathrm{\Lambda}}(-x_{\mathrm{F}}) 
\end{eqnarray}

As the evaluation of the dependence on the transverse momentum
averages the $\mathrm{\Lambda}$ polarization of the forward and backward region
of the $\mathrm{\Lambda}$, the resulting polarization would be zero.
Therefore, by evaluating the dependence on the transverse momenta, the
sign of the polarization for events in the backward region is reversed. The same holds
for the dependence on the invariant masses. 

The measured $\mathrm{\Lambda}$ polarization is shown in fig. \ref {lambdapolarization} as function of the three
kinematic variables.
In this and all following figures the error bars indicate the statistical errors. 
The striking feature of the data is the change of the sign of the
polarization between the two beam momentum settings, which is seen in
all figures. 
In the $\mathrm{\Lambda}$ backward region the $\mathrm{\Lambda}$ polarization is negative for
the beam momentum of 2.70\,GeV/c and positive for the beam momentum of
2.95\,GeV/c. In the forward range the signs are exchanged. This
phenomenon has not been observed before.  Both curves
exhibit within their errors the expected  point symmetry for $\mathrm{cos}\vartheta^{\mathrm{cm}}=0$ and
$x_{\mathrm{F}}=0$. In addition to the difference in sign, the absolute values of the $\mathrm{\Lambda}$ polarization
are significantly different between the two data sets. 

As the variables Feynman~$x_{\mathrm{F}}$ and transverse momentum are not independent,
restricted efficiency in one of these variables will affect the value
of the $\mathrm{\Lambda}$ polarization in the other variable.  For the
transverse $\mathrm{\Lambda}$ momentum the COSY-TOF efficiency (acceptance
$\cdot$ reconstruction efficiency) is nearly constant, the variation
is less than $\pm$~10\%. For the Feynman~$x_{\mathrm{F}}$ variable the
acceptance drops in the range of $|x_{\mathrm{F}}| > 0.8$ from 50\% of the maximum
acceptance to zero at $|x_{\mathrm{F}}| = 1$. But the range of $|x_{\mathrm{F}}| > 0.8$
corresponds to less than 2.8\% of the total phase space volume. Therefore, the small fraction
 of events lying in this range can be neglected for the $p_{\mathrm{t}}$
distribution.

\subsection{Comparison with existing data}
In a first COSY-TOF publication \cite{Bilger1998f} the
$\mathrm{\Lambda}$ polarization was given as a function of the
$\mathrm{\Lambda}$ transversal momentum. But here the change of sign
between the $\mathrm{\Lambda}$ backward and forward region as given in
equation \ref {forwardbackward} was not taken into account.
Therefore, these values cannot be compared with the new results.

\begin{figure}[htpb!]
	\begin {center}
	\includegraphics[width=.5\textwidth]{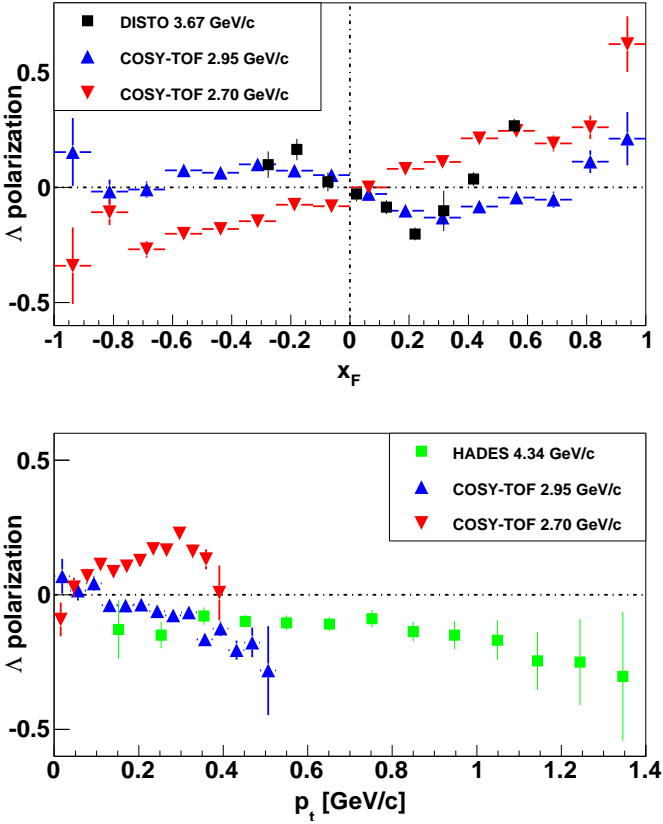}
	\caption{\label{lambda-polarization-vergleich} 
          Upper frame: the $\mathrm{\Lambda}$ polarization as a function of  
          Feynman~$x_{\mathrm{F}}$ in comparison to the results of the DISTO
          experiment.  Bottom frame: the $\mathrm{\Lambda}$
          polarization as a function of the transverse momentum 
          in comparison to the results of the HADES experiment.
          The error bars of the HADES data
          include statistical plus systematic errors.
          The references for the DISTO and HADES data are given in table \ref
          {table1}. }
        \end {center}
\end{figure}

The full kinematic range of the $\mathrm{\Lambda}$ polarization can
only be compared with measurements of DISTO and HADES, which covered a
large fraction of the available phase space.  The results of these
measurements are shown together with the COSY-TOF results in fig. \ref
{lambda-polarization-vergleich}.  The references are given in table
\ref {table1}.  Apart from one data point at $x_{\mathrm{F}}$ $\sim$
0.55 the DISTO data\footnote{ These data are quoted to be preliminary,
  they are scanned from fig. 8 of reference \cite {Choi1998}.}  agree
with the COSY-TOF data of 2.95\,GeV/c.  In dependence on the
transverse momentum the HADES data exhibit a negative
$\mathrm{\Lambda}$ polarization similar to the 2.95\,GeV/c COSY-TOF
results.

\begin{table}[htpb!]
  \centering
  \caption{Denotation of p X\,$\rightarrow$ $\mathrm{\Lambda}$ X measurements given in 
    figs. \ref{lambda-polarization-vergleich},\ref{lambda-polarization-overview}. 
    The reaction types and the proton beam momenta (for collider experiments the invariant masses) are given. }
  \label{table1}
  \begin{tabular}{lcl}

&reference&reaction and \\
&         &beam momentum\\
\hline
    FNAL 1978            & \cite {Heller1978}           &  p+Be \,$\rightarrow$ $ \mathrm{\Lambda}$+X                       \\               
                         &                              & 400.9\,GeV/c                                                      \\
    KEK  1986            & \cite {Abe1986}              &  p+Be (Cu,W)$\rightarrow$ $ \mathrm{\Lambda}$+X                   \\                  
                         &                              & 12.9\,GeV/c                                                       \\
    CERN ISR 1987        & \cite {Smith1987c}           &  p+p \,$\rightarrow$ $\mathrm{\Lambda}$+X                         \\                        
                         &                              & $\sqrt{s}$ = $\numrange{3}{61}$\,GeV                              \\
    AGS 1988 13.5\,GeV/c & \cite {Bonner1988}           &  p+Be\,$\rightarrow$ $\mathrm{\Lambda}$+X                         \\                     
                         &                              & 13.5\,GeV/c                                                       \\
    AGS 1988 18.5\,GeV/c & \cite {Bonner1988}           &  p+Be\,$\rightarrow$ $\mathrm{\Lambda}$+X                         \\                          
                         &                              & 18.5\,GeV/c                                                       \\
    AGS 1996             & \cite {Felix1996,Felix1999a} &  p+p \,$\rightarrow$ pK$^{+}$$\mathrm{\Lambda}$ N($\pi^{+}\pi^{-}$) \\
                         &                              & 27.5\,GeV/c                                                       \\
    FNAL 1989            & \cite {Lundberg1989}         &  p+Be (Cu,Pb)\,$\rightarrow$ $\mathrm{\Lambda}$+X                 \\                   
                         &                              & 400.9\,GeV/c                                                      \\
    FNAL 1991            & \cite {Felix2002}            &  p+p \,$\rightarrow$ pK$^{+}$$\mathrm{\Lambda}$                    \\                     
                         &                              & 800.9\,GeV/c                                                      \\
    FNAL 1994            & \cite {Ramberg1994}          &  p+Be\,$\rightarrow$ $\mathrm{\Lambda}$+X                         \\                          
                         &                              & 800.9\,GeV/c                                                      \\
    DISTO 1998           & \cite {Choi1998}             &  p+p \,$\rightarrow$ pK$^{+}$$\mathrm{\Lambda}$                    \\                          
                         &                              & 3.67\,GeV/c                                                       \\
    CERN NA48 1999       & \cite {Fanti1999}            &  p+Be \,$\rightarrow$ $\mathrm{\Lambda}$+X                        \\                          
                         &                              & $\sqrt{s}$ = 61\,GeV                                              \\
    HERA-B 2006          & \cite {Abt2006}              &  p+C(W)\,$\rightarrow$ $\mathrm{\Lambda}$+X                       \\                         
                         &                              & 920.9\,GeV/c                                                      \\
    HADES 2014           & \cite {Agakishiev2014}       &  p+Nb\,$\rightarrow$ $\mathrm{\Lambda}$+X                         \\                         
                         &                              & 4.34\,GeV/c                                                       \\
    ATLAS 2015           & \cite {Aad2015}              &  p+p\,$\rightarrow$ $\mathrm{\Lambda}$+X                          \\                          
&& $\sqrt{s}$ = 7\,TeV \\
  \end{tabular}
\end{table}

\begin{figure}[hbtp]
	\begin {center}
	\includegraphics[width=.5\textwidth]{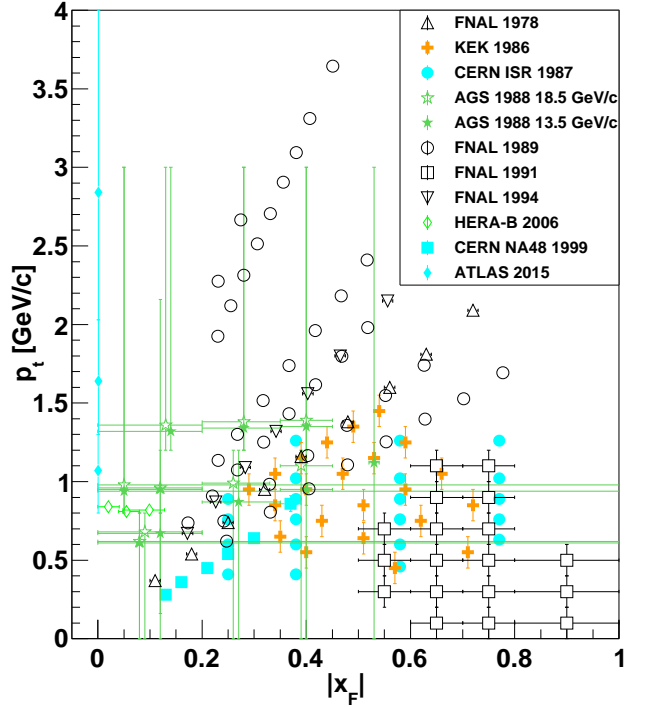}
	\caption{\label{lambda-polarization-overview} Kinematic
          regions of $\mathrm{\Lambda}$ polarization measurements, for which the
          dependence of the $\mathrm{\Lambda}$ polarization on the transverse
          momentum and on Feynman~$x_{\mathrm{F}}$ are explicitly given. The bin width of
          the measurements are indicated by error bars. The references
          for the measurements are given in table \ref {table1}.
        }  
        \end {center}
\end{figure}

As there are conjectures in the literature (see for instance \cite
{Neal2006}) that the $\mathrm{\Lambda}$ polarization may be
independent of the beam momentum and target material, the COSY-TOF
results are compared with measurements with beam momenta ranging from
near threshold up to 1\,TeV/c and with measurements on different
target materials.  These experiments cover only a small part of the
available phase space.  The range of $x_{\mathrm{F}}$ and
$p_{\mathrm{t}}$, for which the $\mathrm{\Lambda}$ polarization is
specified, is plotted by lines in
fig. \ref{lambda-polarization-overview}. The labels of the different
experiments in the legends of
figs. \ref{lambda-polarization-vergleich},\ref{lambda-polarization-overview}
are referenced in table \ref {table1}.

Most of these measurements are inclusive, and thus the separation of
directly produced $\mathrm{\Lambda}$ from $\mathrm{\Sigma}^0
\rightarrow \mathrm{\gamma} \mathrm{\Lambda}$ decay is not
possible. The first determination of the $\mathrm{\Sigma}^0$
polarization \cite {Dukes1987} yields a value of $\approx 30\%$ with
opposite sign compared to the $\mathrm{\Lambda}$
polarization. Therefore, the inclusive measurements are expected to
reveal a $\mathrm{\Lambda}$ polarization that is lower than the
polarization of the directly produced $\mathrm{\Lambda}$.  The ratio
of directly produced $\mathrm{\Lambda}$ to those from the
$\mathrm{\Sigma}^0$ decay is dependent on the beam momentum and may be
dependent on the kinematical regions of the measurements. The first
exclusive measurement of the $\mathrm{\Lambda}$ polarization of the
reaction pp\,$\rightarrow$ pK$^{+}$$\mathrm{\Lambda}$ at CERN \cite
{Henkes1992} yielded values up to $60\%$, which are significantly
larger than the results obtained by the inclusive measurements. The
high polarization values were confirmed by an exclusive measurement at
FNAL \cite {Felix2002}.

\begin{figure}[hbtp]    
	\begin {center}
	\includegraphics[width=.5\textwidth]{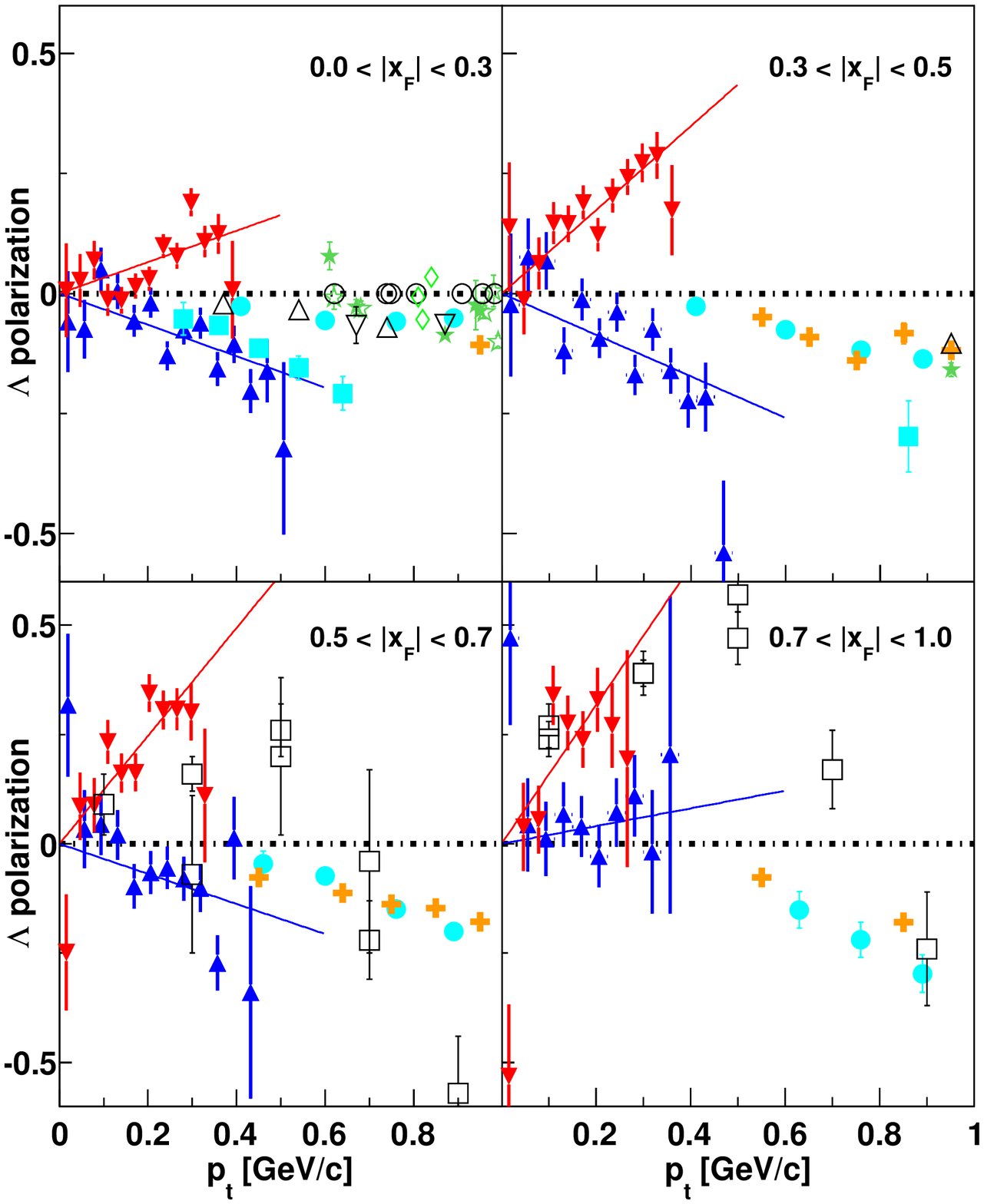}
	\caption{\label{lambdapol-xfl} Comparison of the COSY-TOF data
          (blue, filled triangles up for 2.95\,GeV/c and red, filled
          triangles down for 2.70\,GeV/c) with data from the literature, the
          symbols are the same as in 
          fig. \ref{lambda-polarization-overview}.  For better visibility 
          error bars of the abscissa are not plotted. The
          range of Feynman~$x_{\mathrm{F}}$ is indicated in the plots. The
          COSY-TOF data have been fit with a straight
          line, the fit parameters are given in table \ref {table2}.
        }  
        \end {center}
\end{figure}

In order to incorporate the restricted ranges of the measurements
given in fig. \ref{lambda-polarization-overview}, the results of
COSY-TOF  are
recalculated for 4 ranges in $x_{\mathrm{F}}$ and compared with values of the literature
that are inside these ranges (see fig. \ref{lambdapol-xfl}).

For each interval in $x_{\mathrm{F}}$ a linear fit is applied to the COSY-TOF
data. The results are given in table \ref {table2}.  The slope
of the 2.70\,GeV/c data fit is positive for all $x_{\mathrm{F}}$ intervals and
rises with $x_{\mathrm{F}}$.  The slope of the 2.95\,GeV/c data fit is
negative and within the errors the same in the first three $x_{\mathrm{F}}$
intervals.  For the extreme $\mathrm{\Lambda}$ forward and backward range
($|x_{F}|>$0.7) it is compatible with zero. No systematic agreement
with the data from the literature can be detected.  Nearly all
measurements of the literature show a negative $\mathrm{\Lambda}$ polarization.
Apart from the data of one experiment \cite
{Felix2002} below $p_{\mathrm{t}}$ = 0.6\,GeV/c  the COSY-TOF 2.70\,GeV/c data are the only 
ones with a  positive polarization for all intervals.

\begin{table}[htbp]
  \centering
  \caption{The results from the linear fit in fig. \ref {lambdapol-xfl} 
    \newline ($P_{\mathrm{N}} = a_{1} \cdot p_{\mathrm{t}} [GeV/c]$)}
  \label{table2}
  \begin{tabular}{ccr}
data range &  \multicolumn{2}{c} {$a_{1}$ [(GeV/c)$^{-1}$] }           \\
  $|x_{\mathrm{F}}|$ & $p_{\mathrm{b}}$ = 2.70\,GeV/c                    & $p_{\mathrm{b}}$ = 2.95\,GeV/c \\
\hline
$\numrange{0.0}{0.3}$  & $0.33 \pm 0.04$                         & $-0.33\pm 0.04$ \\
$\numrange{0.3}{0.5}$  & $0.87 \pm 0.06$                         & $-0.43\pm 0.05$ \\
$\numrange{0.5}{0.7}$  & $1.23 \pm 0.08$                         & $-0.34\pm 0.07$ \\
$\numrange{0.7}{1.0}$  & $1.59 \pm 0.18$                         & $ 0.20\pm 0.15$ \\
  \end{tabular}
\end{table}

\begin{figure}[hbtp]
	\begin {center}
	\includegraphics[width=.5\textwidth]{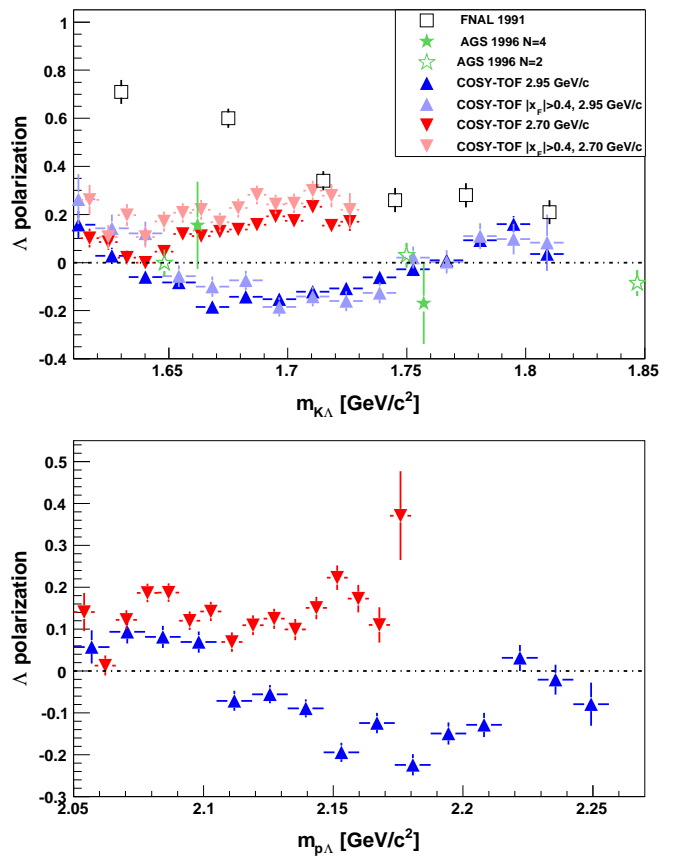}
	\caption{\label{lambdapol-mkl} Top: comparison of the  $\mathrm{\Lambda}$ polarization
          as a function of the K$\mathrm{\Lambda}$ invariant mass. In order to show the effect of 
          the restriction of the data to the range of Feynman~$x_{\mathrm{F}}$ as in the literature data,
          the COSY-TOF data are plotted with and without this restriction. In the legend
          of the AGS data N gives the number of $\mathrm{\pi}^{+}\mathrm{\pi}^{-}$ pairs which are evaluated
          together with pK$\mathrm{\Lambda}$. Bottom: COSY-TOF data of the $\mathrm{\Lambda}$ polarization 
          as a function of the  p$\mathrm{\Lambda}$ invariant mass.
        }  
        \end {center}
\end{figure}

The dependence of the $\mathrm{\Lambda}$ polarization on the kaon-$\mathrm{\Lambda}$
invariant mass is given by the experiments BNL E766 and FNAL
E690  \cite {Felix2002,Felix2006,Felix2006a}.
These data are compared with the data of COSY-TOF in
fig. \ref {lambdapol-mkl}. Because the data of BNL and FNAL are restricted
to Feynman~$x_{\mathrm{F}}$ region of $|x_{\mathrm{F}}|>0.4$, we apply the same restriction
to the COSY-TOF data. The BNL data set varies by including
$\mathrm{\pi}^{+}\mathrm{\pi}^{-}$ pairs to the pK$\mathrm{\Lambda}$ final state.  Only the
data with 2 and 4 pions are shown in fig. \ref {lambdapol-mkl}.
No obvious similarities between these data sets and the COSY-TOF data can be
found. For the sake of completeness  the $\mathrm{\Lambda}$ polarization is plotted as a function of  the p$\mathrm{\Lambda}$
 invariant mass in the lower part of fig.  \ref
{lambdapol-mkl}. No data in the literature are existing for comparison.

\subsection{Dependence of the $\mathrm{\Lambda}$ polarization on the cm momentum}

\begin{figure}[hbtp]
	\begin {center}
	\includegraphics[width=.5\textwidth]{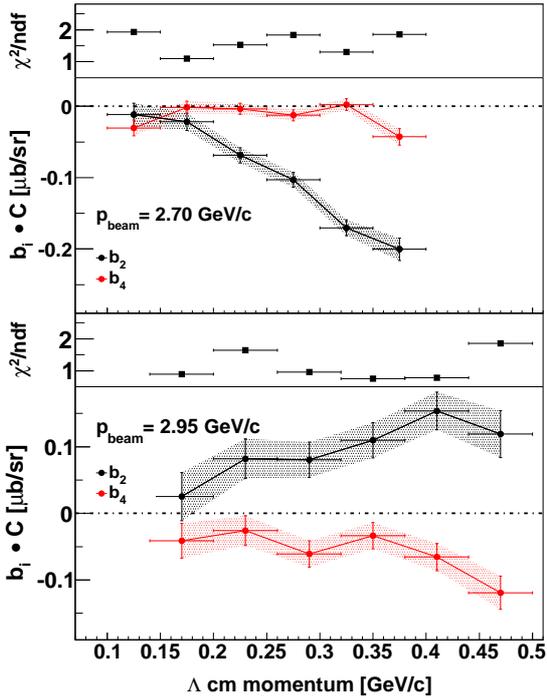}
	\caption{\label{lambdaPolarization_pcmBins_Coefficients}
          The variation of the Legendre coefficients $b_2$ (black) and $b_4$
          (red) as defined in eq. \ref {legendrecoe} with the cm
          $\mathrm{\Lambda}$ momentum.  The upper panel contains  the
          data of 2.70\,GeV/c beam momentum, the lower one contains the 2.95\,GeV/c 
          data. Above each plot  the reduced
          $\chi^2$ of the fit is plotted.
        }  
        \end {center}
\end{figure}

In order to characterize the variation of the $\mathrm{\Lambda}$ polarization with
$\mathrm{cos}\vartheta_{\mathrm{\Lambda}}^{\mathrm{cm}}$ by only two parameters, the
$\Lambda$ polarization is multiplied by the differential cross section and fitted with the sum of the associated Legendre polynomials
$L_{2}^{1}$ and $L_{4}^{1}$:
\begin {equation} \label{legendrecoe}
     P_{\mathrm{N}}(\mathrm{cos}\vartheta_{\mathrm{\Lambda}}^{\mathrm{cm}}) \cdot \frac{\mathrm{d}\sigma_{0}}{\mathrm{d}\Omega} = b{_2} L_{2}^{1}(\mathrm{cos}\vartheta_{\mathrm{\Lambda}}^{\mathrm{cm}}) +b{_4} L_{4}^{1}(\mathrm{cos}\vartheta_{\mathrm{\Lambda}}^{\mathrm{cm}})
\end {equation}
d$\sigma_{0}/\mathrm{d}\Omega$ is the spin averaged differential cross section determined as
described in sec. \ref{AN_LEGENDRE}. 
 These polynomials are chosen as they have the required roots and
 point symmetry. $L_{2}^{1}$ describes the data with a structure
 of one maximum/minimum each in the forward and backward region, while the polynomial
 $L_{4}^{1}$ has two maxima/minima in each region.  The data are
 divided into six bins of the $\mathrm{\Lambda}$ cm momentum as shown
 by the  error bars at the abscissa in fig. \ref {lambdaPolarization_pcmBins_Coefficients}
 and fit according to eq. \ref {legendrecoe}. The fit results
 do not improve by adding the $l=6$ Legendre polynomial, which would
 add structures with three maxima/minima in each region.  The
 variation of the coefficients $b_{2}$ and $b_{4}$ with the cm
 momentum are shown in fig. \ref {lambdaPolarization_pcmBins_Coefficients} 
 for both beam  momenta.
 
For the 2.70\,GeV/c data the coefficient $b_{4}$ is within the error bars constant and close to zero, 
while the magnitude of the coefficient $b_{2}$ increases 
nearly linearly with the $\mathrm{\Lambda}$ cm  momentum. For the 2.95\,GeV/c data the magnitudes
of both coefficients rise with the momentum. The distribution and fit for
each momentum bin is shown in the appendix A in figs. 
\ref {lambdaPolarization_pcmBins_270.eps}, \ref {lambdaPolarization_pcmBins_295.eps}.   

\section{Analyzing power $A_{\mathrm{N}}$}
\subsection{Results }
The analyzing power $A_{\mathrm{N}}$ describes  the left-right
asymmetry  of the final state particles induced by the vertical beam polarization.
The analyzing power is determined by applying the  equation: 
\begin{equation}\label{eq:analyzingpower}
A_{\mathrm{N}} = \frac{1} {P_{\mathrm{B}}\cdot \mathrm{cos}(\varphi)} \cdot  \frac{N^{\uparrow}(\varphi) - N^{\downarrow}(\varphi)}{N^{\uparrow}(\varphi) + N^{\downarrow}(\varphi)}
\end{equation}
$\varphi$ is the azimuth angle,
$N$ denotes the count rates,  
and $\uparrow$, $\downarrow$ indicate the direction of the beam
polarization. $P_{\mathrm{B}}$ is the absolute value of the beam polarization.
Three beam proton analyzing powers can be determined: $A_{\mathrm{N}}(\varphi^{\mathrm{p}})$, 
 $A_{\mathrm{N}}(\varphi^{\mathrm{K}})$, and $A_{\mathrm{N}}(\varphi^{\mathrm{\Lambda}})$ 
depending on  which final state particle is considered.

In contrast to the $\mathrm{\Lambda}$ polarization the symmetry of the initial
system is broken because the beam proton is polarized. Therefore, no
symmetry in the functional dependence on the scattering angle and on the
Feynman~$x_{\mathrm{F}}$ are expected. The following boundary conditions are
given:  $A_{\mathrm{N}}(|\mathrm{cos}(\vartheta^{\mathrm{cm}})|=1)=0$ and $A_{\mathrm{N}}(|x_{\mathrm{F}}|=1)=0$.
As the minimum of the transverse momentum corresponds to $|\mathrm{cos}(\vartheta^{\mathrm{cm}})|=1$, the
analyzing power at $p_{\mathrm{t}}$ = 0\,GeV/c must also be zero.
  
\begin{figure}[hbtp]
	\begin {center}
	\includegraphics[width=.5\textwidth]{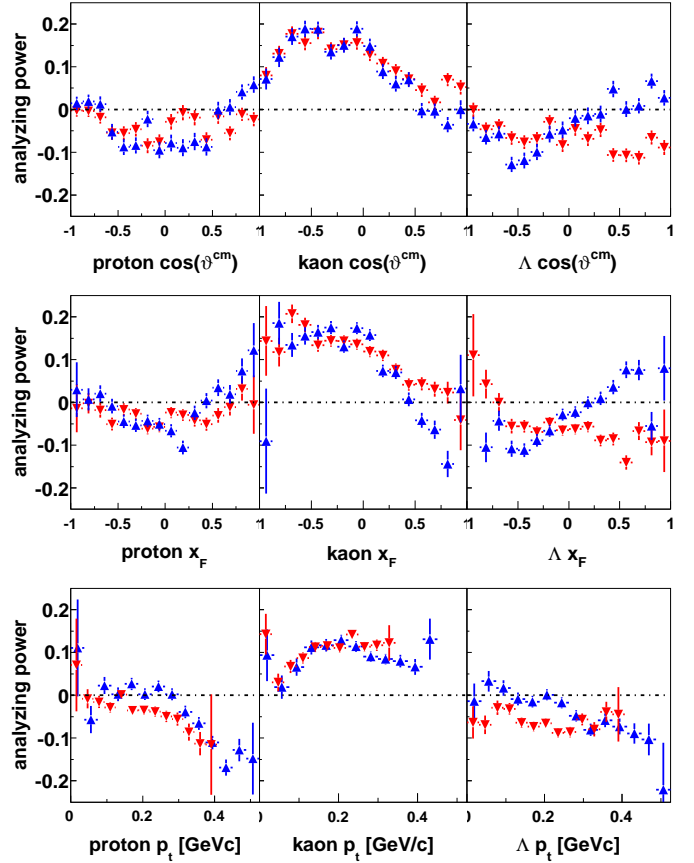}
	\caption{\label{analyzing_power_measurement} 
          The analyzing power is shown for two beam momenta: 2.70\,GeV/c
          (red triangles down) and 2.95\,GeV/c (blue triangles up). Top: dependence on
          the cm scattering angle. Middle: dependence on the
          Feynman~$x_{\mathrm{F}}$. Bottom: dependence on the transverse momentum.
        }  
        \end {center}
\end{figure}

The results of the  analyzing power measured by the proton, kaon, and $\mathrm{\Lambda}$ asymmetries 
are shown in fig. \ref {analyzing_power_measurement}. The
analyzing powers are given as a function of the cm scattering angle, the
Feynman~$x_{\mathrm{F}}$, and the transverse momentum of the corresponding final state particle.

While the analyzing power generating the  p and K$^{+}$ asymmetries does not change significantly with the 
beam momentum, larger differences are found for the analyzing power from  the $\mathrm{\Lambda}$ asymmetry. 
The analyzing power of the proton and of the $\mathrm{\Lambda}$ asymmetry are mainly
negative, while the analyzing power of the kaon asymmetry is essentially
positive. For transverse momenta below 0.25\,GeV/c the analyzing
power of the proton asymmetry is nearly zero and decreases linearly to
$-0.2$ above this momentum. The analyzing power of the kaon asymmetry shows an inverse behavior: for
momenta below 0.25\,GeV/c it increases nearly linearly from 0 to 0.10 
and  it is constant above this momentum.

\subsection{Comparison with existing data }
\begin{figure}[htpb!]
	\begin {center}
	\includegraphics[width=.5\textwidth]{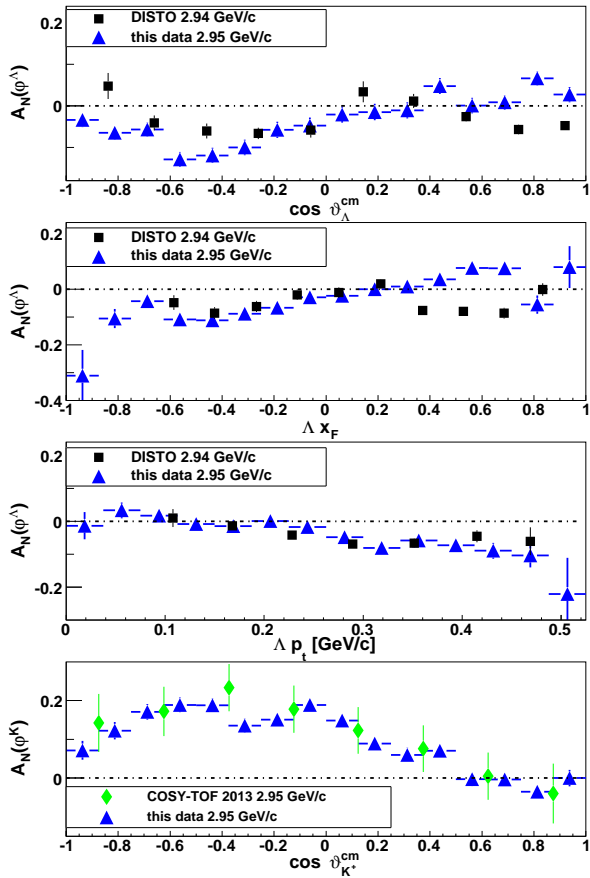}
	\caption{\label{ay-vergleich}In the top panels the
          analyzing power measured at 2.95\,GeV/c is compared with data from DISTO,
          which were measured at 2.94\,GeV/c \cite {Maggiora2001}. The
          bottom panel compares the analyzing power of the kaon asymmetry at 2.95
         \,GeV/c with a previous analysis including only 20\% of the COSY-TOF data \cite {Roeder2013}.
        }  
        \end {center}
\end{figure}

The analyzing power  has only
been measured exclusively for the associated strangeness production by DISTO and COSY-TOF.  The
DISTO data were measured at beam momenta of 3.67\,GeV/c, 3.31\,GeV/c,
and 2.94\,GeV/c \cite {Maggiora2001,Balestra2001}.  The
 analyzing power determined with the $\mathrm{\Lambda}$  asymmetry of the 2.94\,GeV/c data\footnote{ The data
  are scanned from fig. 4 of reference \cite {Maggiora2001}, they are
  quoted to be preliminary.}  are compared in fig. \ref {ay-vergleich}
with the COSY-TOF data of 2.95\,GeV/c as a function of the
scattering angle,  Feynman~$x_{\mathrm{F}}$, and the transverse
momentum. While the dependence on the transverse momentum of both
data sets exhibits a similar behavior, discrepancies are seen in $\mathrm{cos}\vartheta^{\mathrm{cm}}$ and $x_{\mathrm{F}}$: 
for the forward range the
DISTO data yield a negative analyzing power, the COSY-TOF data have positive values.

 A subset of the COSY-TOF data, which had been measured in 2010,  was
 evaluated in \cite {Roeder2013} for the determination of the spin
 triplet p$\mathrm{\Lambda}$ scattering length. The comparison of the 
 analyzing power of the kaon asymmetry  is given
 in the last plot of fig. \ref {ay-vergleich}. Both analyses exhibit
 consistent characteristics of this analyzing power.


\subsection{Description with Legendre polynomials:  } \label{AN_LEGENDRE}

\begin{figure}[t]
	\begin {center}
	\includegraphics[width=.5\textwidth]{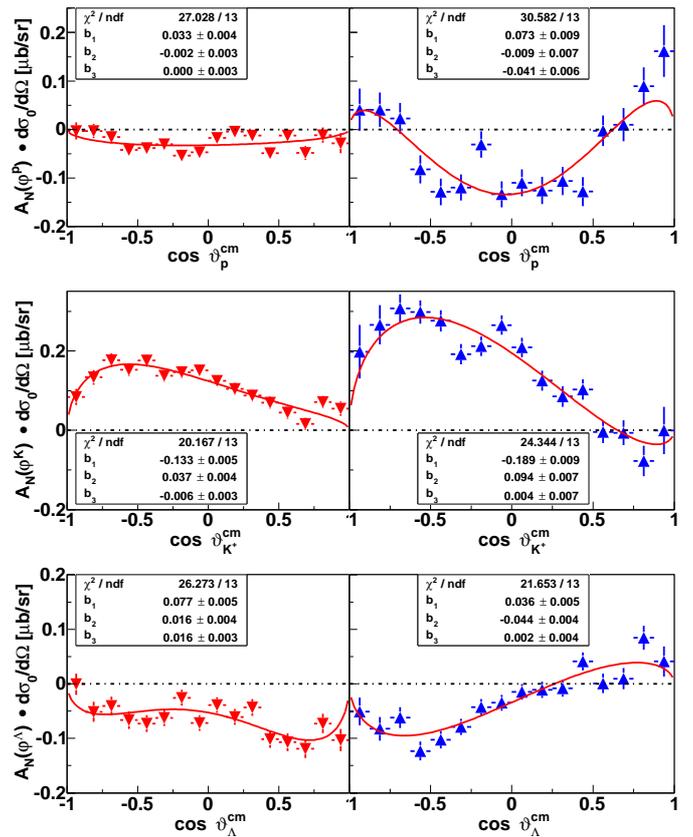}
	\caption{\label{polynomial_fits} The product of the analyzing
          power and the unpolarized cross section
          $A_{\mathrm{N}}\cdot\mathrm{d}\sigma_{0}/\mathrm{d}\Omega$ for the three final state particles
          is shown by the triangles, left column for 2.70\,GeV/c beam
          momentum, right column for 2.95\,GeV/c beam momentum. The fit
          results for the sum of the Legendre polynomials are shown by
          the red lines. The values of the coefficients $b_{\mathrm{i}}$ are
          given in the plots.
        }  
        \end {center}
\end{figure}

The differential cross  section  of  proton-proton interactions with a beam polarization $P_{\mathrm{B}}$ in the $\pm \mathrm{y}$ 
direction and an unpolarized target is given by (\cite {Saha1983,Niskanen1981}):

\begin {equation} \label {differential_cross_section}
\frac{\mathrm{d}\sigma(\mathrm{cos}\vartheta^{*})}{\mathrm{d}\Omega} = \frac{\mathrm{d}\sigma_{0}(\mathrm{cos}\vartheta^{*})}{\mathrm{d}\Omega} + P_{\mathrm{B}} \frac{\mathrm{d}\sigma_{\mathrm{y}}(\mathrm{cos}\vartheta^{*})}{\mathrm{d}\Omega}
\end {equation}
d$\sigma_{0}(\mathrm{cos}\vartheta^{*})$/${\mathrm{d}\Omega}$ is the spin averaged differential cross section and d$\sigma_{y}(\mathrm{cos}\vartheta^{*})$/${\mathrm{d}\Omega}$ is  the  spin dependent cross section. 
The analyzing power $A_{\mathrm{N}}$ is given by
\begin {equation} \label {ay}
 A_{\mathrm{N}} = \frac{ \mathrm{d} \sigma_{\mathrm{y}}(\mathrm{cos}\vartheta^{*})/\mathrm{d}\Omega}{  \mathrm{d} \sigma_{0}(\mathrm{cos}\vartheta^{*})/\mathrm{d}\Omega} 
\end {equation}
The differential cross sections  can be expressed by a sum of Legendre polynomials:
\begin {eqnarray} \label {analyzing_power}
 \mathrm{d}\sigma_{0}(\mathrm{cos}\vartheta^{*})/\mathrm{d}\Omega &=& \frac{1}{4\pi}\sum_{\mathrm{n}}a_{\mathrm{n}}L^{0}_{\mathrm{n}}\nonumber\\
 \mathrm{d}\sigma_{y}(\mathrm{cos}\vartheta^{*})/\mathrm{d}\Omega &=& \frac{1}{4\pi}\sum_{\mathrm{n}}b_{\mathrm{n}}L^{1}_{\mathrm{n}}
\end {eqnarray}
$L^{0}_{\mathrm{n}}$ ($L^{1}_{\mathrm{n}}$) are the associated Legendre functions of
order 0 (1). $a_{\mathrm{n}}$ are the related coefficients of order 0, $\mathrm{n} = 0,1,2 \dots$
and $b_{\mathrm{n}}, \mathrm{n} =1,2,3 \dots $ are the related coefficients of order~1.
In order to determine the coefficients $b_{\mathrm{n}}$, the product of the
analyzing power and the unpolarized cross section
$A_{\mathrm{N}} \cdot\mathrm{d}\sigma_{0}/\mathrm{d}\Omega$ is fit with the  $\mathrm{n}=1,2,3$ associated Legendre
functions. The unpolarized cross section d$\sigma_{0}/\mathrm{d}\Omega$ is
determined by fitting the angular distribution (see  fig. 3 in reference \cite
{Jowzaee2016}) with associated Legendre functions of order 0 up
to the degree of $\mathrm{n}=2$.  The fit results of $A_\mathrm{N}\cdot\mathrm{d}\sigma_0/\mathrm{d}\Omega$ are
shown in fig. \ref {polynomial_fits}.

In contrast to proton-proton reactions with two particles in the final
state -- e.g. pp\,$\rightarrow$ d$\pi^+$ -- no direct conclusions on the
partial waves can be drawn from the composition of the Legendre
coefficients.

\subsection{Dependence of the Legendre Coefficients on the particle momenta }

\begin{figure}[t]
      \begin {center}
      \includegraphics[width=.5\textwidth]{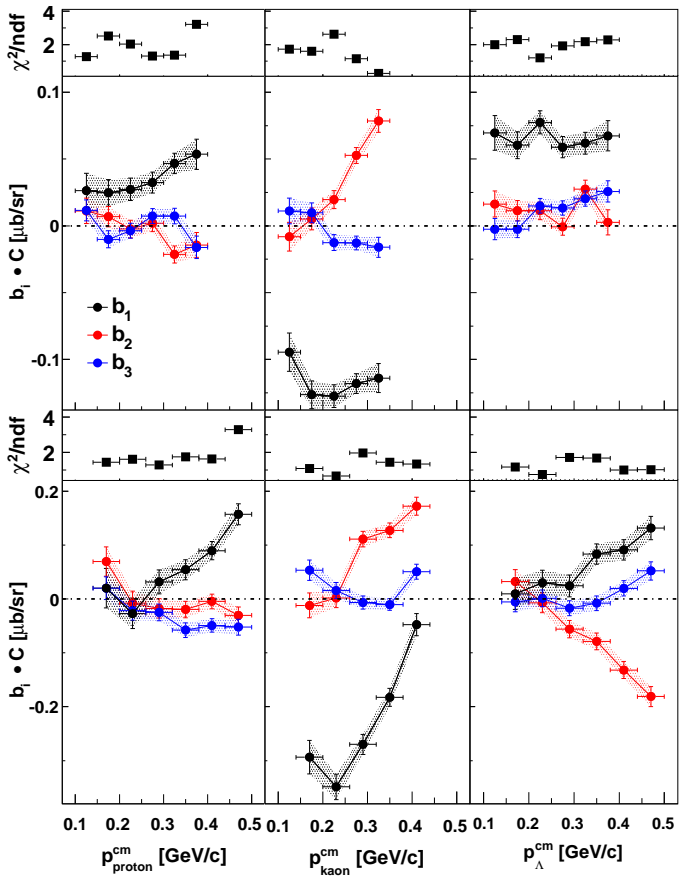}
      \caption{\label{legendre-coefficients-pdependence} 
         Legendre coefficients obtained
         by fitting the analyzing power plotted as a function of the 
         cm momenta. The coefficients are distinguished
         by the color code: black: $b_1$, red: $b_2$, and blue:
         $b_3$.  The values are plotted for 2.70\,GeV/c (top row) 
         and 2.95\,GeV/c (bottom row).
        The reduced $\chi^2$ values of the fits are shown above each plot.
       }  
       \end {center}
\end{figure}

In order to determine the dependence of the Legendre coefficients on
the particle cm momentum, the data are divided into six bins of the
particle momentum. The analyzing power
distribution is calculated for each bin and fit with the Legendre polynomials
$L^{1}_{\mathrm{n}} (\mathrm{n}=1,2,3)$. 
The distributions of the analyzing power and the fit in each bin
are shown in the appendix A in figs. \ref{AnalyzingPower_Coeff_pcmBins_Proton270}
-  \ref{AnalyzingPower_Coeff_pcmBins_Lambda295}. The range of the momentum of
each bin is indicated in the plots.  
The momentum dependence of the  coefficients is shown in
fig.~\ref{legendre-coefficients-pdependence}.

By omitting the coefficients $b_2$, $b_3$ or both in the fit and by
comparing the reduced $\chi^2$ of these fits it is observed that
below cm momenta of $\numrange{200}{300}$\,MeV/c the analyzing power can be
described with the first Legendre polynomial alone. Above this
momentum the analyzing power of the proton asymmetry needs the
third Legendre polynomial and the analyzing powers of the $\mathrm{\Lambda}$ and
kaon asymmetry need  the second Legendre polynomial.
  
The spin triplet part of the p-$\mathrm{\Lambda}$ scattering length
can be derived from the kaon cm momentum distribution and the
analyzing power \cite {Gasparyan2005}.  This method is based on
the variation of the coefficient $b_{1}$ (see eq. \ref
{analyzing_power}) with the kaon cm momentum in the range of the
highest kaon momenta (fig. \ref {legendre-coefficients-pdependence},
second column). At 2.95 GeV/c (lower row of fig. \ref
{legendre-coefficients-pdependence}) the coefficient $b_{1}$ is for the highest kaon momenta close
to zero ($-0.05 \pm 0.02$), thus the spin
triplet part of the $\mathrm{\Lambda}$-p scattering length can not be
determined with the required precision \cite {Roeder2013}.  But for
the 2.70\,GeV/c data this coefficient is $-0.12 \pm 0.01$ for the
higher kaon momenta (upper row of fig. \ref
{legendre-coefficients-pdependence}), enabling this determination.
The results of the scattering length from the 2.70 GeV/c data are discussed in a separate paper
\cite {Hauenstein2015}.

\section{Spin transfer coefficient: $D_{\mathrm{NN}}$}

\subsection{Results}
The transfer of the beam-proton polarization to the hyperon is
quantified by the spin transfer coefficient $D_{\mathrm{NN}}$.  The common
definition (see for instance \cite {Bonner1988}) implies that it is
positive if the polarization of the hyperon is aligned with the beam
polarization, negative if the hyperon polarization is oriented opposite to the
beam polarization, and zero if the hyperon polarization is independent
of the beam polarization.

The spin transfer coefficient is calculated with the following formula:
\begin{equation}\label{eq:dnn5}
D_{\mathrm{NN}}(\vartheta^*) = \frac{4} {\alpha P_{\mathrm{B}}} \epsilon_D (\vartheta^*)
\end{equation}
$\alpha$ is the $\mathrm{\Lambda}$ decay asymmetry parameter,
$P_{\mathrm{B}}$ the magnitude
of the beam polarization, and $\epsilon_D (\vartheta^*)$ is calculated
from the differences of count rates depending on the orientation of
the beam polarization, the $\mathrm{\Lambda}$ polarization, and the hemisphere
of the detector. For a detailed discussion see ref.~\cite {Hauenstein2016}.

\begin{figure}[htpb!]
	\begin {center}
	\includegraphics[width=.5\textwidth]{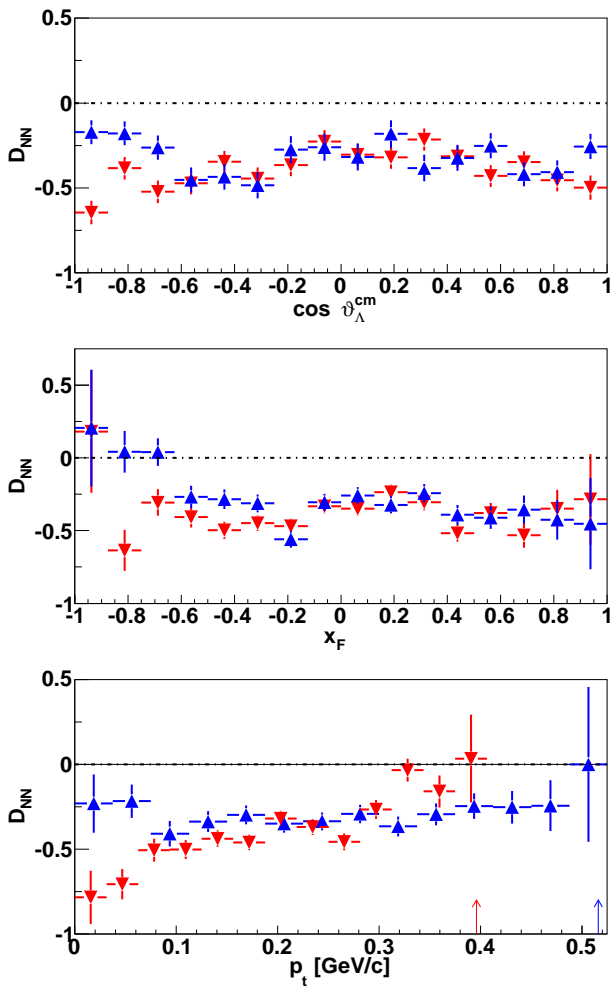}
	\caption{\label{DNN} 
          Spin transfer coefficient for 2.70\,GeV/c
          (red triangles down) and 2.95\,GeV/c (blue triangles up). Top: dependence on
          the $\mathrm{\Lambda}$ cm scattering angle. Middle: dependence on the $\mathrm{\Lambda}$
          Feynman~$x_{\mathrm{F}}$. Bottom: dependence on the $\mathrm{\Lambda}$ transverse momentum.
          For both beam momenta 
          the maxima
          of the transverse momenta are indicated with arrows in the
          lower part of the plot.
        }  
        \end {center}
\end{figure}

The spin transfer coefficient $D_{\mathrm{NN}}$  is shown in fig. \ref {DNN} as a function of
 the $\mathrm{\Lambda}$ scattering angle, Feynman~$x_{\mathrm{F}}$, and the 
transverse momentum of the  $\mathrm{\Lambda}$. 
The measurements show that the coefficient $D_{\mathrm{NN}}$ is
negative for nearly all regions, thus the hyperon polarization is
opposite to the beam polarization.  It is mainly constant as a
function of the scattering angle and of Feynman~$x_{\mathrm{F}}$
within the interval of $x_{\mathrm{F}}$= $[-0.4, +1.0]$.  For the
2.70\,GeV/c data $D_{\mathrm{NN}}$ increases linearly with the
transverse momentum with a slope of (1.3 $\pm$0.2) (GeV/c)$^{-1}$
starting with $-0.6$ at $p_{\mathrm{t}}$ = 0\,GeV/c.  This behavior is
not repeated for the data of 2.95\,GeV/c beam momentum, where the fit
gives a gradient consistent with zero.

\subsection{Comparison with existing data}

\begin{figure}[t]
	\begin {center}
	\includegraphics[width=.5\textwidth]{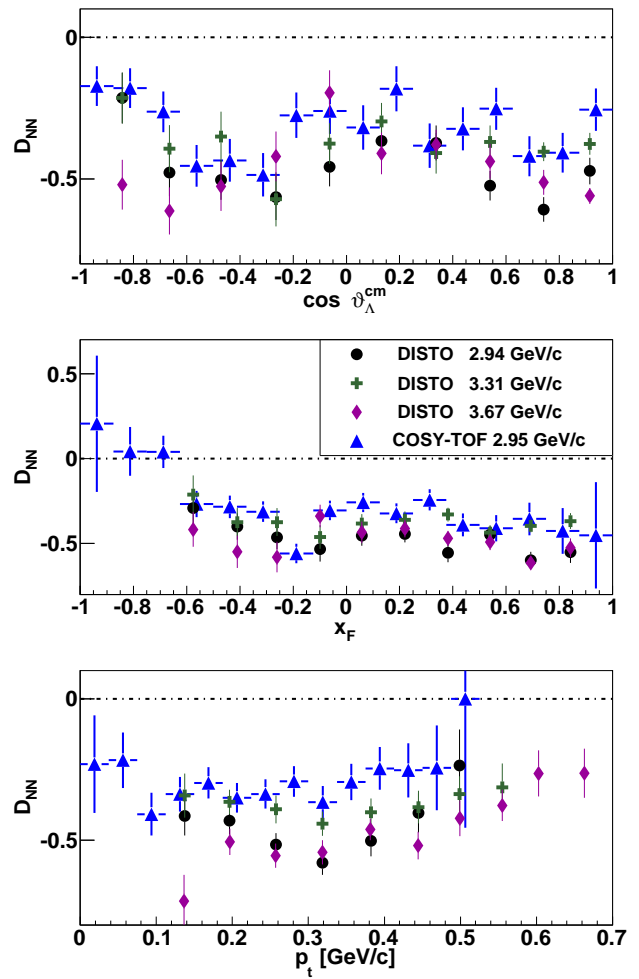}
	\caption{\label{dnn_vergleich} 
          The spin transfer coefficient is shown for the
          DISTO data \cite {Maggiora2001} and the 2.95\,GeV/c COSY-TOF
          data. For a better clarity, the 2.70\,GeV/c
          data of COSY-TOF are omitted.  Top:
          dependence on the $\mathrm{\Lambda}$ cm scattering angle. Middle:
          dependence on the $\mathrm{\Lambda}$ Feynman~$x_{\mathrm{F}}$. Bottom: dependence on
          the $\mathrm{\Lambda}$ transverse momentum.
        }  
        \end {center}
\end{figure}

The first measurement of the spin transfer coefficient of pp
\,$\rightarrow$ $\mathrm{\Lambda}$ X was performed in 1975 at a beam momentum of
6\,GeV/c \cite {Lesnik1975}.  $D_{\mathrm{NN}}$ was measured to be $\approx 0$ for Feynman~$x_{\mathrm{F}} <
0.6$  and for $x_{\mathrm{F}}> 0.6$ to be $\approx -0.5$.
The range of the corresponding transverse momentum is not given in the
paper and the contamination with $\mathrm{\Lambda}$'s of the $\mathrm{\Sigma}^{0}$ decay
is not discussed. In 1988 $D_{\mathrm{NN}}$ was measured at beam momenta of
13.3\,GeV/c and 18.5\,GeV/c with a beryllium target \cite
{Bonner1988}. The measurements covered  transverse momenta from 0 to 3
GeV/c  with Feynman~$x_{\mathrm{F}}$ mainly in the forward range. The ratio of
$\mathrm{\Lambda}$ from the $\mathrm{\Sigma}^{0}$ decay to direct production was
estimated to be 40\%. No
significant deviations of $D_{\mathrm{NN}}$ from zero were found. A third
measurement with a beam momentum of 200\,GeV/c and a liquid hydrogen
target covered the forward range of $x_{\mathrm{F}}$ \cite {Bravar1997}. The
fraction of $\mathrm{\Lambda}$'s from $\mathrm{\Sigma}^{0}$ decay is not discussed in
the paper. $D_{\mathrm{NN}}$ was
measured to be compatible with zero for 0.1\,GeV/c $<$ $p_{\mathrm{t}}$ $<$ 0.6\,GeV/c, and for 0.6\,GeV/c $<$ $p_{\mathrm{t}}$ $<$
1.5\,GeV/c $D_{\mathrm{NN}}$ is positive and in the range of $\numrange{0.3}{0.4}$. 
The results of these measurements are not in
agreement with the present data and they exhibit no consistent
dependence on Feynman~$x_{\mathrm{F}}$. Therefore, it cannot be determined how
strongly the polarization transfer depends on the beam momentum. The
first exclusive measurements of the polarization transfer were
performed by the DISTO experiment \cite {Maggiora2001}. Here pK$\mathrm{\Lambda}$
and pK$\mathrm{\Sigma}^{0}$ events are identified in the pK missing mass
spectrum. For the beam momentum of 3.67\,GeV/c the admixture of $\mathrm{\Lambda}$ from $\mathrm{\Sigma}^0$ decay was
measured to be  lower than 30\% \cite {Balestra1999a}. 
These data are compared with the COSY-TOF results of $p_{\mathrm{b}}$ = 2.95\,GeV/c
 in fig. \ref {dnn_vergleich}.  
Both data sets exhibit a negative $D_{\mathrm{NN}}$, which is approximately constant.
The magnitude of $D_{\mathrm{NN}}$ measured by  COSY-TOF is smaller compared with the DISTO data. The mean value of 
the COSY-TOF data in the   $x_{\mathrm{F}}$ range of $[-0.5, +1.0]$ is $-0.37 \pm 0.02$,
the corresponding mean value of the DISTO data is $-0.46 \pm 0.01$. The mean values of each measurement
are given in table \ref {dnn_table}.

\begin{table}[htbp]
  \centering
  \caption{ Mean values of  $D_{\mathrm{NN}}$ for  $x_{\mathrm{F}}$ = $[-0.5, +1.0]$ }
  \label{dnn_table}
  \begin{tabular}{lc}
data set& $\overline{D_{\mathrm{NN}}}$      \\
 
\hline
DISTO 2.94 GeV/c & $-0.50 \pm 0.02$     \\
DISTO 3.31 GeV/c & $-0.38 \pm 0.02$     \\
DISTO 3.67 GeV/c & $-0.50 \pm 0.02$     \\
\hdashline
COSY-TOF 2.70 GeV/c & $-0.39 \pm 0.02$    \\
COSY-TOF 2.95 GeV/c & $-0.34 \pm 0.02$    \\
  \end{tabular}
\end{table}

  The indication 
that $D_{\mathrm{NN}}$ changes to positive values for $x_{\mathrm{F}} < -0.6$ cannot
be confirmed by the DISTO data, as the acceptance of this detector for
that range is too small \cite {Balestra1999}.

\subsection{Comparison with model calculations}

\begin{figure}[htpb!]
	\begin {center}
	\includegraphics[width=.4\textwidth]{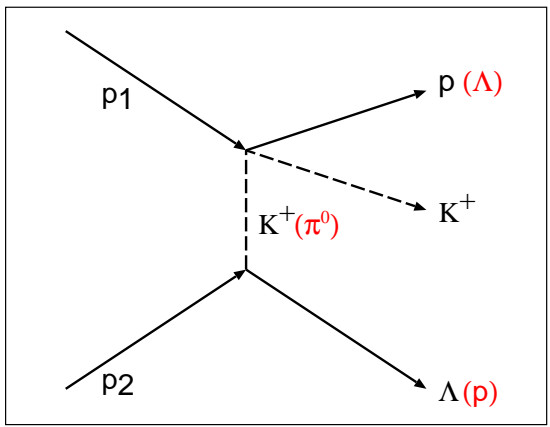}
	\caption{\label{feynmangraph}Feynman graph of the associated
          strangeness production, with either kaon exchange or pion
          exchange (red symbols in brackets). p$_{1,2}$ are the beam
          and target proton, their order can be exchanged.
        }  
        \end {center}
\end{figure}

As developed by Laget \cite
{Laget1991} the dependence of the spin transfer coefficient on the
Feynman~$x_{\mathrm{F}}$ can be characterized by assuming pion or kaon exchange
mechanism for the associated strangeness production. For pion exchange
the upper vertex in the Feynman diagram of fig. \ref{feynmangraph} is
p + $\mathrm{\pi}$\,$\rightarrow$ K$^{+}$ + $\mathrm{\Lambda}$.  
Both sides of this vertex  have negative parity, if angular momenta are 0.
For a polarized beam proton, the $\mathrm{\Lambda}$
spin will have the same direction as the proton spin ($D_{\mathrm{NN}}$ = +1).
In case of a kaon exchange the lower vertex in the Feynman diagram is
given by p\,$\rightarrow$ K$^{+}$ + $\mathrm{\Lambda}$. Here the parity
changes, and therefore, the $\mathrm{\Lambda}$ spin has to be opposite compared to
the beam proton polarization direction ($D_{\mathrm{NN}} = -1$).
For backward $\mathrm{\Lambda}$ regions the $\mathrm{\Lambda}$ is essentially 
produced on the unpolarized target proton. Thus, in  this region    $D_{\mathrm{NN}}$  is expected to be zero.
 
As for $x_{\mathrm{F}} > 0$  the data exhibit $D_{\mathrm{NN}}\approx
-0.3$, the conclusion from the Laget model indicates the existence of
a mixture of kaon and pion exchange mechanism. The backward region
exhibits the same value for $D_{\mathrm{NN}}$, only for
$x_{\mathrm{F}} < -0.6$ the 2.95\,GeV/c data give values of
$D_{\mathrm{NN}}$ which are compatible with zero.  A later
calculation \cite {Faldt2005} considers in addition the exchange of a
${\mathrm\rho}$ meson. It is shown that $D_{\mathrm{NN}}$ can have
negative values without assuming a kaon exchange. $D_{\mathrm{NN}}$
is given as a function of the ratio of the ${\mathrm\pi}$ exchange
amplitude ($D_{\mathrm{\pi}}$) to the ${\mathrm\rho}$ exchange
amplitude ($B_{\mathrm{\rho}}$).  For $D_{\mathrm{NN}} = -0.3$ there
are two solutions: $D_{\mathrm{\pi}}/B_{\mathrm\rho} =
0.3$ and $D_{\mathrm{\pi}}/B_{\mathrm\rho} = 1.5 $ (fig. 3 of
ref. \cite {Faldt2005}).  In addition, this calculation gives the ratio of the
total cross sections $R$ = (pn $\rightarrow$ nK$^{+}$$\mathrm{\Lambda}$) / (pp
$\rightarrow$ pK$^{+}$$\mathrm{\Lambda}$) as a function of 
$D_{\mathrm{\pi}}/B_{\mathrm\rho}$. The value of $D_{\mathrm{NN}} = -0.3$ implies the ratio
$R$ = 4 and $R$ = 6. 
This ratio will be examined by a future publication,
where results from  COSY-TOF of pn $\rightarrow$ pK$^{0}$$\mathrm{\Lambda}$
at a beam momentum of 2.95~GeV/c will be presented.

\section{Summary}
The $\mathrm{\Lambda}$ polarization, the analyzing power determined by
the asymmetry of the final state particles, and the $\mathrm{\Lambda}$
spin transfer coefficient are measured exclusively in the reaction
pp\,$\rightarrow$ pK$^{+}\mathrm{\Lambda}$ at beam momenta of
2.70\,GeV/c and 2.95\,GeV/c.

It is shown that the $\mathrm{\Lambda}$ polarization changes
significantly in its magnitude and sign with the beam momentum. This
is the first time that this effect is observed.  In contrast to all
existing data the 2.70 GeV/c COSY-TOF data exhibit a positive
$\mathrm{\Lambda}$ polarization in forward direction.  The 2.95 GeV/c
data are in agreement with the measurements of DISTO and HADES. 
Comparisons of the $\mathrm{\Lambda}$ polarization of both beam
momenta with data from high beam momentum experiments do not yield similar
characteristics.

In contrast to the $\mathrm{\Lambda}$ polarization the analyzing power
does not change significantly with the beam momentum, only the
analyzing power measured with the $\mathrm{\Lambda}$ asymmetry differs
for these beam momenta. The results at 2.95 GeV/c beam momentum are in
agreement with a measurement of DISTO at a beam momentum of 2.94
GeV/c. Apart from the DISTO data no other data  exist for 
comparison.  The  dependence of the Legendre
coefficients on the particle cm momenta yield different behavior
between the two beam momenta, especially for the analyzing power
measured with the kaon asymmetry.

The measurement of the spin transfer coefficient $D_{\mathrm{NN}}$
yields a negative value of about $-0.3$. Apart from the backward region
($x_{\mathrm{F}} < -0.6$) this value does not significantly vary with
the beam momentum. The comparison with data from the DISTO experiment
yields an agreement of the sign, but the magnitude of
$D_{\mathrm{NN}}$ measured with DISTO is about 25\% larger compared
with the COSY-TOF data.  A model calculation, which compares
$\mathrm{\rho}$ and $\mathrm{\pi}$ exchange, connects the value of
$D_{\mathrm{NN}}$ with the ratio of the cross sections 
associated strangeness production in pp and pn reactions. The measured value of
$D_{\mathrm{NN}}$ implies that this ratio is 4 or 6.  This ratio will be determined with existing COSY-TOF data.

\begin{acknowledgement}
The authors want to thank the COSY crew for the excellent beam preparation,
J. Uehlemann and N. Paul for the operation of the demanding LH$_2$ target. 
This work was supported by grants from Forschungszentrum J\"{u}lich (COSY-FFE),
by the European Union Seventh Framework program (FP7/2007-2013) under
grant agreement 283286, and by the Foundation for Polish Science through the MPD programme.
Helpful comments of C. Wilkin are gratefully acknowledged. 
\end{acknowledgement}
\clearpage
\section {Appendix A}
This appendix contains the individual fits of the observables as a function of the cm angle, which are separated into bins of the cm momentum.
Inside each figure the fit results and the range of the cm momentum are given.
\begin{figure*}[p]
	\centering
	\includegraphics[width=.95\textwidth]{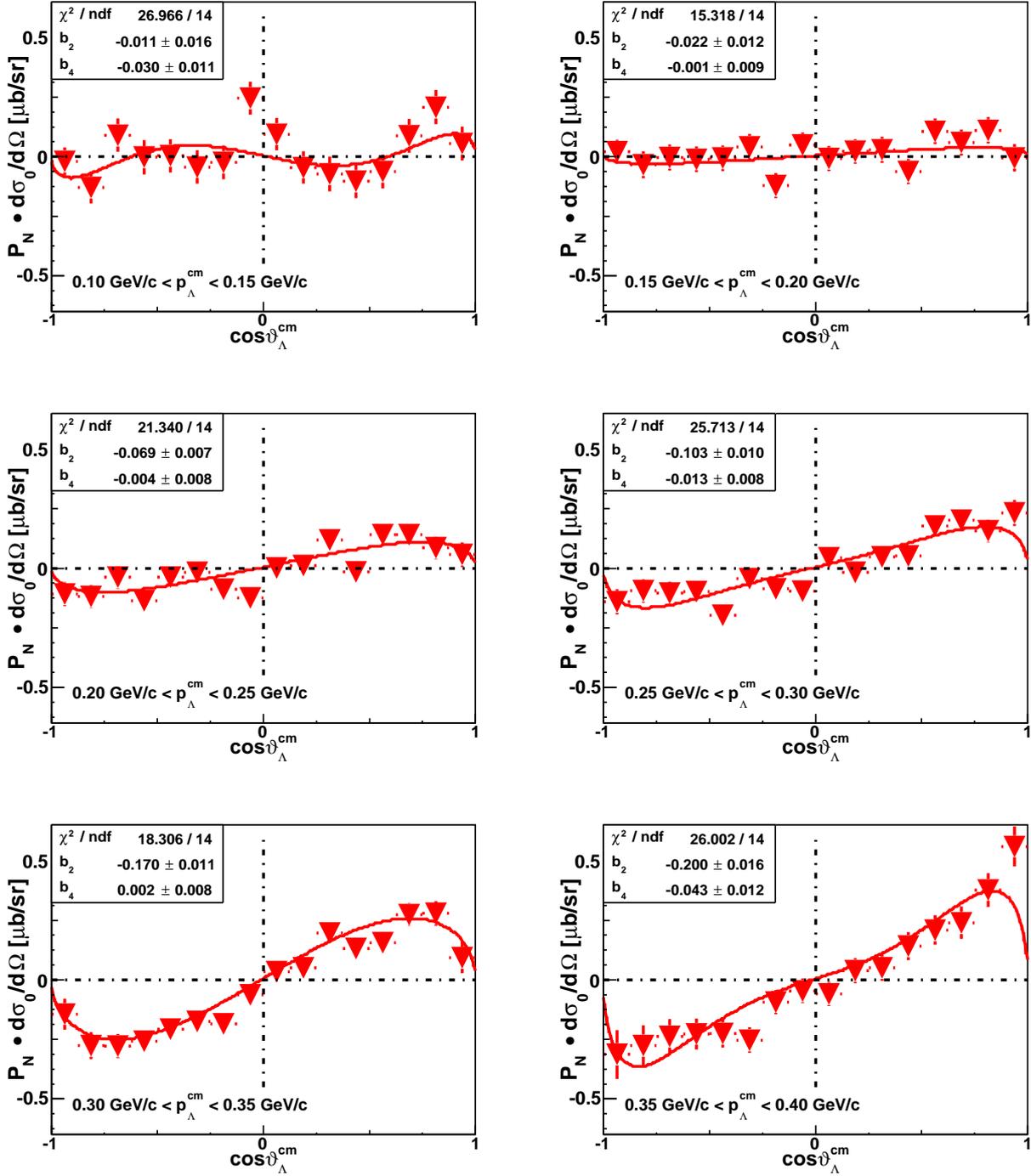}
	\caption{\label{lambdaPolarization_pcmBins_270.eps} The $\mathrm{\Lambda}$ polarization multiplied with the differential cross section of the 2.70\,GeV/c data is shown for different ranges
          of the  $\mathrm{\Lambda}$ cm momentum. The distributions have been fit with Legendre polynomials according to 
          eq. \ref{legendrecoe}. The fit results and the limits of the $\mathrm{\Lambda}$ cm momenta are given in each 
          panel.
        }
\end{figure*}\clearpage
\begin{figure*}[p]
	\centering
	\includegraphics[width=.95\textwidth]{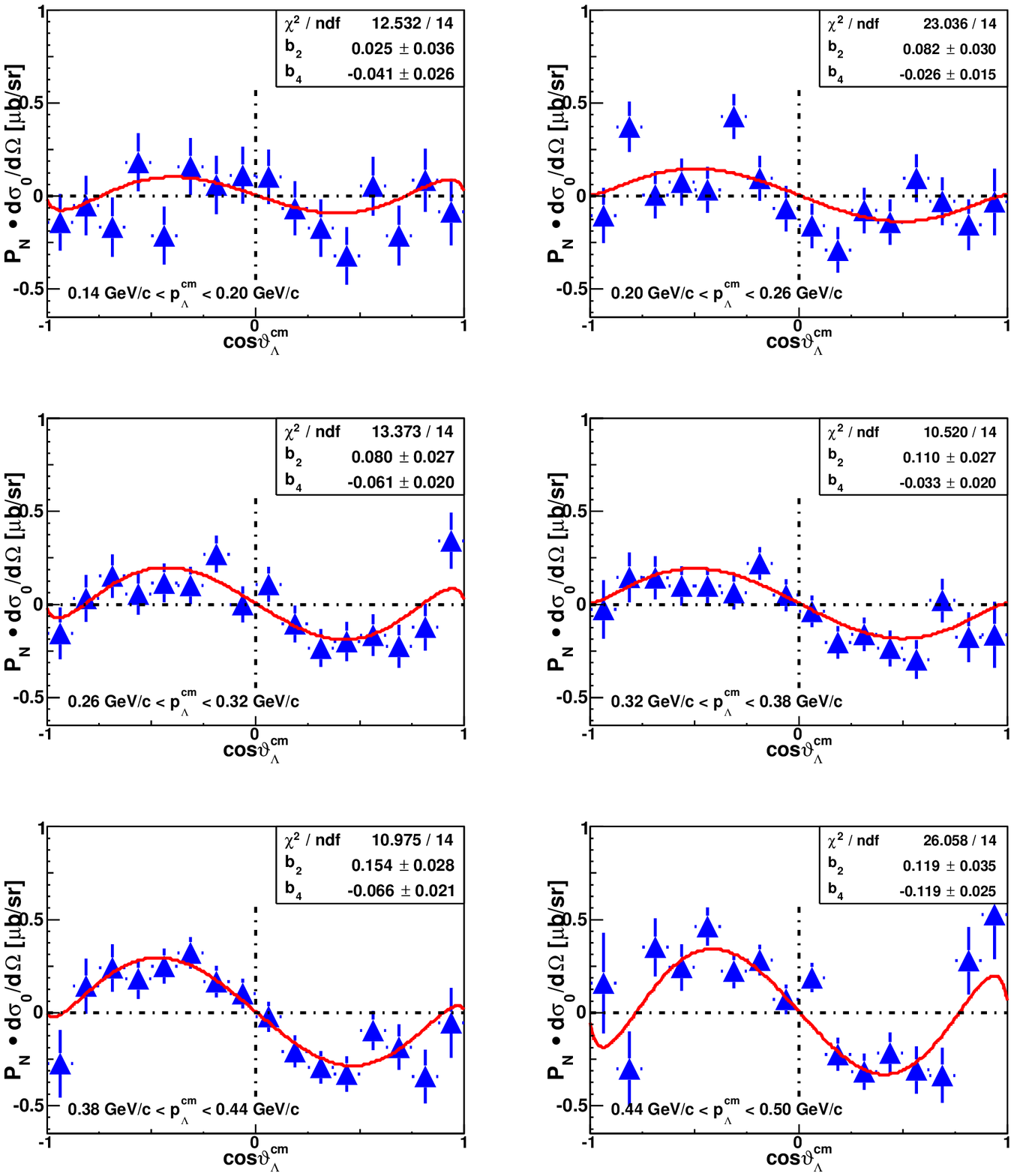}
	\caption{\label{lambdaPolarization_pcmBins_295.eps} The $\mathrm{\Lambda}$ polarization multiplied with the differential cross section of the 2.95\,GeV/c data is shown for different ranges
          of the  $\mathrm{\Lambda}$ cm momentum. The distributions have been fit with Legendre polynomials according to 
          eq. \ref{legendrecoe}. The fit results and the limits of the $\mathrm{\Lambda}$ cm momenta are given in each 
          panel. 
        }
\end{figure*}\clearpage
\begin{figure*}[p]
	\centering
	\includegraphics[width=.95\textwidth]{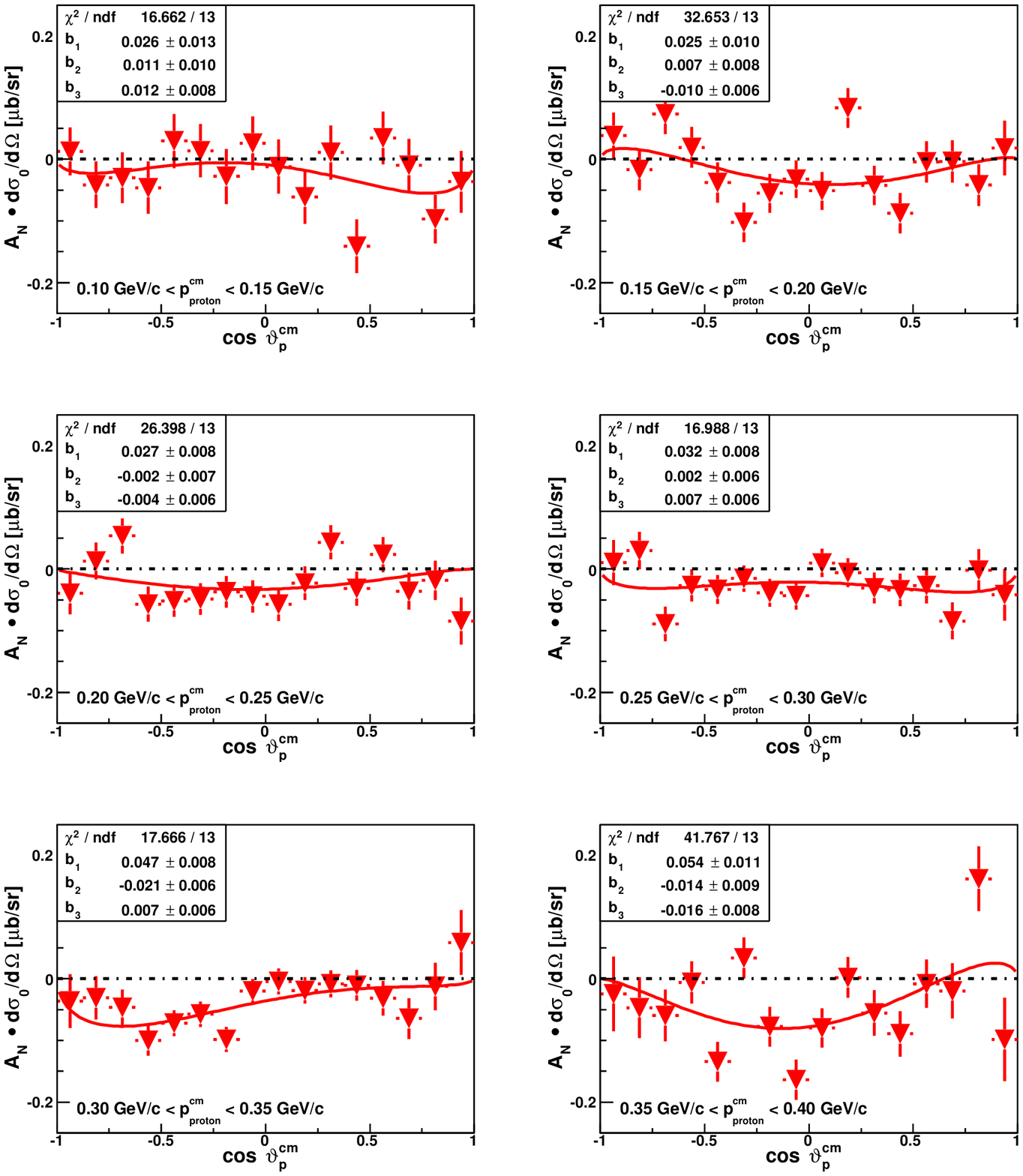}
	\caption{\label{AnalyzingPower_Coeff_pcmBins_Proton270}  The proton analyzing power multiplied with the differential cross section  is shown for different ranges
          of the  proton cm momentum for the 2.70\,GeV/c data. The distributions have been fit with Legendre polynomials according to 
          eq. \ref{analyzing_power}. The fit results and the limits of the proton cm momenta are given in each 
          panel.
        }
\end{figure*}\clearpage
\begin{figure*}[p]
	\centering
	\includegraphics[width=.95\textwidth]{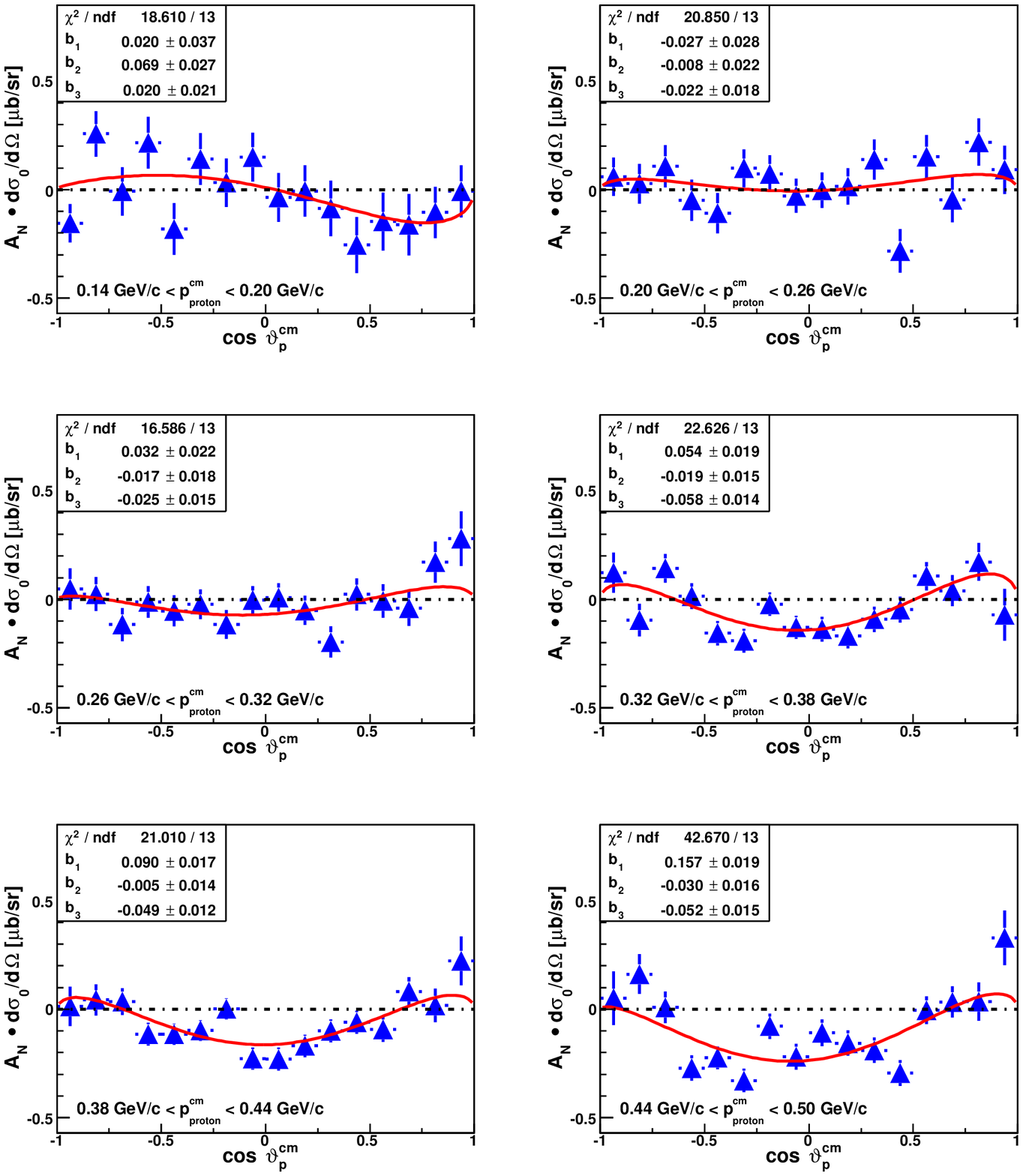}
	\caption{\label{AnalyzingPower_Coeff_pcmBins_Proton295} The proton analyzing power multiplied with the differential cross section  is shown for different ranges
          of the  proton cm momentum for the 2.95\,GeV/c data. The distributions have been fit with Legendre polynomials according to 
          eq. \ref{analyzing_power}. The fit results and the limits of the proton cm momenta are given in each 
          panel.
        }
\end{figure*}\clearpage
\begin{figure*}[p]
	\centering
	\includegraphics[width=.95\textwidth]{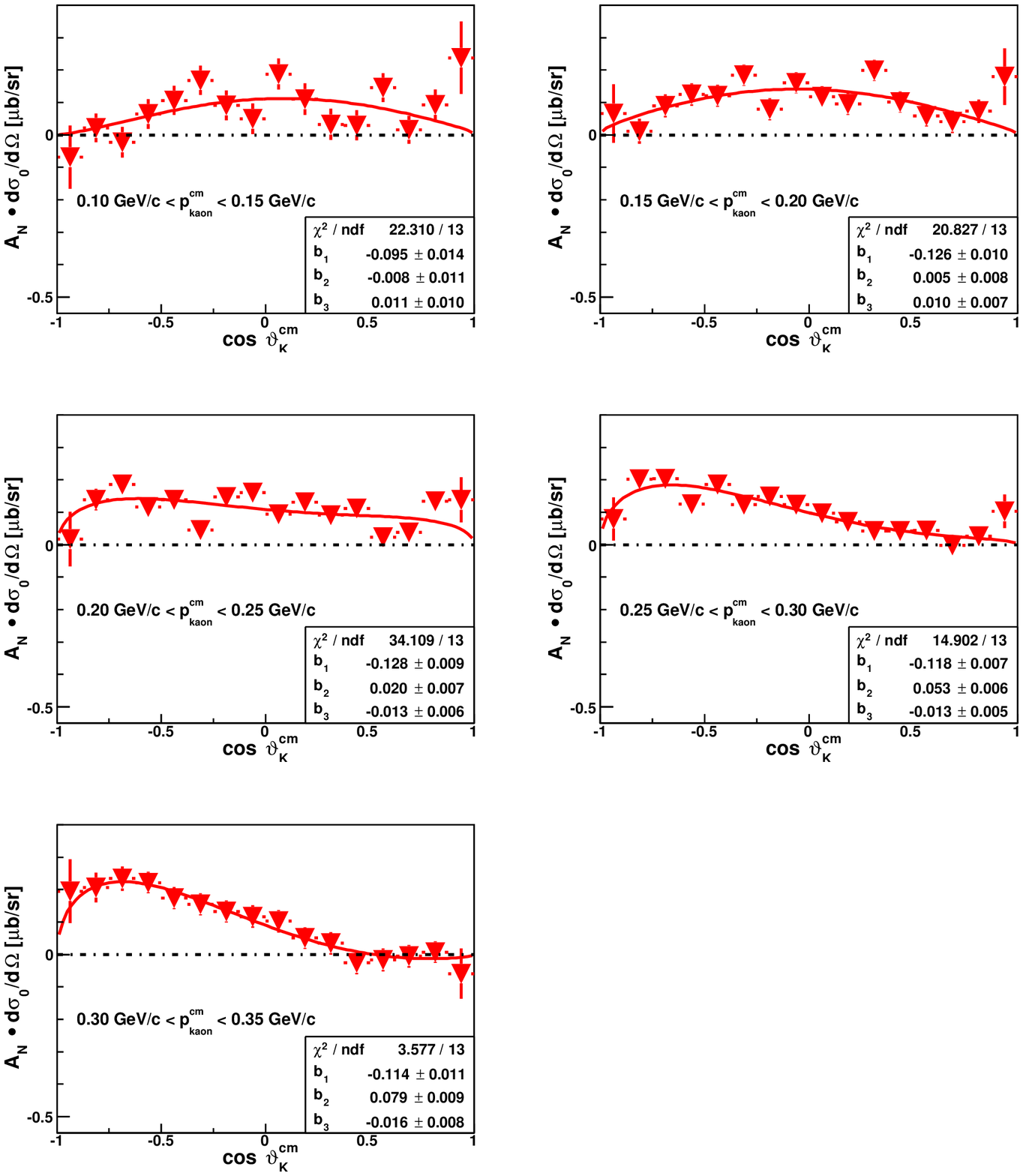}
	\caption{\label{AnalyzingPower_Coeff_pcmBins_Kaon270}  The kaon analyzing power multiplied with the differential cross section  is shown for different ranges
          of the  kaon cm momentum for the 2.70\,GeV/c data. The distributions have been fit with Legendre polynomials according to 
          eq. \ref{analyzing_power}. The fit results and the limits of the kaon cm momenta are given in each 
          panel.
        }
\end{figure*}\clearpage
\begin{figure*}[p]
	\centering
	\includegraphics[width=.95\textwidth]{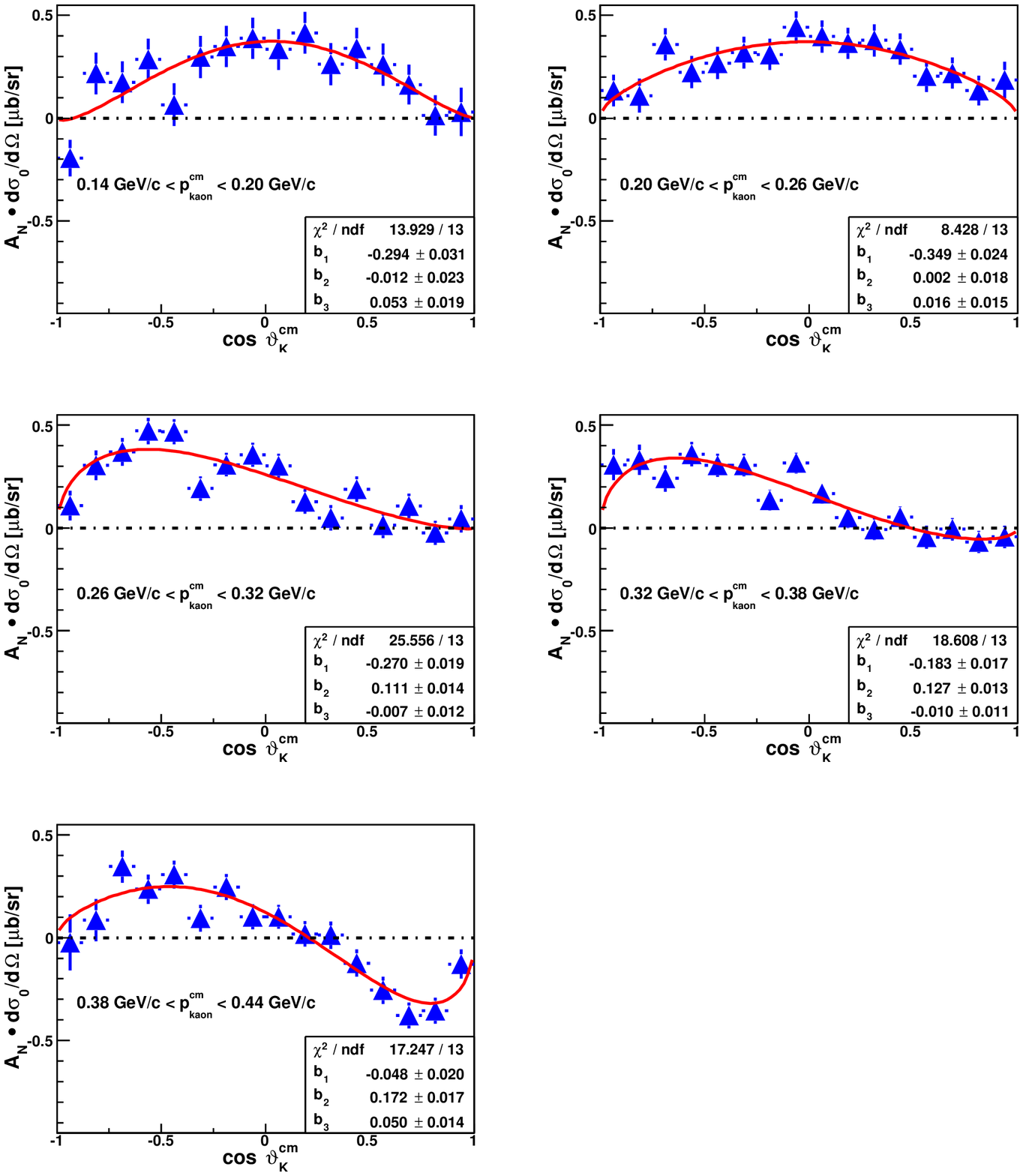}
	\caption{\label{AnalyzingPower_Coeff_pcmBins_Kaon295} The kaon analyzing power multiplied with the differential cross section  is shown for different ranges
          of the  kaon cm momentum for the 2.95\,GeV/c data. The distributions have been fit with Legendre polynomials according to 
          eq. \ref{analyzing_power}. The fit results and the limits of the kaon cm momenta are given in each 
          panel.
        }
\end{figure*}\clearpage
\begin{figure*}[p]
	\centering
	\includegraphics[width=.95\textwidth]{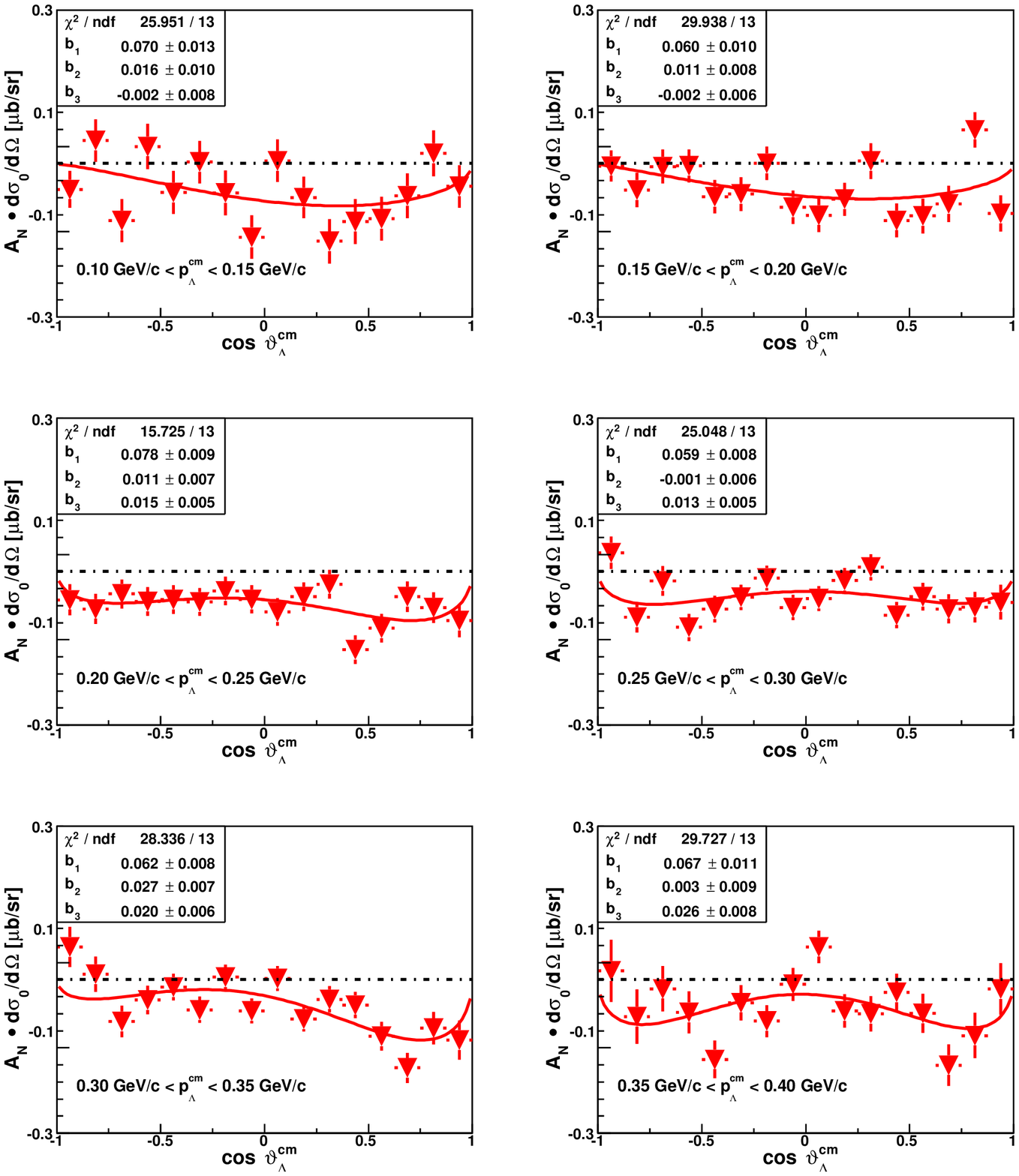}
	\caption{\label{AnalyzingPower_Coeff_pcmBins_Lambda270}  The $\mathrm{\Lambda}$ analyzing power multiplied with the differential cross section  is shown for different ranges
          of the  $\mathrm{\Lambda}$ cm momentum for the 2.70\,GeV/c data. The distributions have been fit with Legendre polynomials according to 
          eq. \ref{analyzing_power}. The fit results and the limits of the $\mathrm{\Lambda}$ cm momenta are given in each 
          panel.
        }
\end{figure*}\clearpage
\begin{figure*}[p]
	\centering
	\includegraphics[width=.95\textwidth]{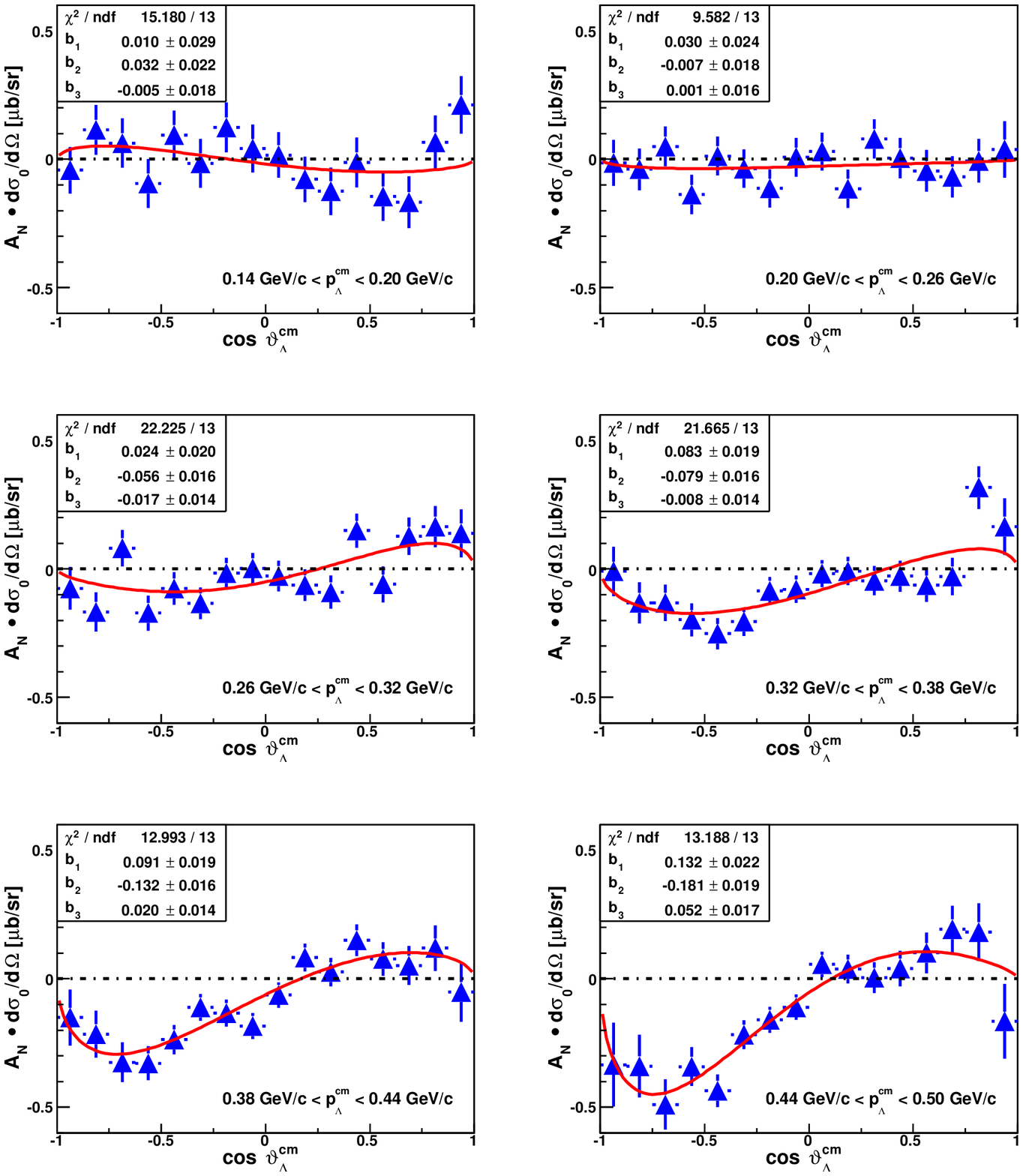}
	\caption{\label{AnalyzingPower_Coeff_pcmBins_Lambda295}The $\mathrm{\Lambda}$ analyzing power multiplied with the differential cross section  is shown for different ranges
          of the  $\mathrm{\Lambda}$ cm momentum for the 2.95\,GeV/c data. The distributions have been fit with Legendre polynomials according to 
          eq. \ref{analyzing_power}. The fit results and the limits of the $\mathrm{\Lambda}$ cm momenta are given in each 
          panel. 
        }
\end{figure*}\clearpage

\section {Appendix B}
This appendix contains data tables with the measured quantities corresponding to 
figs.  \ref {lambdapolarization}, \ref {analyzing_power_measurement}, and \ref {DNN}.
\clearpage
\begin{table*}
  \centering
  \caption{$\mathrm{\Lambda}$ polarization, dependence on $\mathrm{\Lambda}$ cm scattering angle }
  \label{tablePol1}
  \begin{tabular}{c|cc|cc}
    &\multicolumn{2} {c|}{p$_{\mathrm{b}}=2.70 $\,GeV/c}  &\multicolumn{2} {c}{ p$_{\mathrm{b}}=2.95 $\,GeV/c }  \\
    cos${\mathrm{\vartheta}_{\mathrm{\Lambda}}^{*}}$                       & polarization                   & $\pm$&polarization                    &  $\pm$    \\
    \hline 
    -0.9375 & -0.128 & 0.028 & -0.023 & 0.026  \\ 
    -0.8125 & -0.172 & 0.027 & 0.026 & 0.026  \\ 
    -0.6875 & -0.130 & 0.026 & 0.060 & 0.026  \\ 
    -0.5625 & -0.161 & 0.026 & 0.081 & 0.027  \\ 
    -0.4375 & -0.184 & 0.026 & 0.098 & 0.028  \\ 
    -0.3125 & -0.100 & 0.026 & 0.139 & 0.028  \\ 
    -0.1875 & -0.142 & 0.026 & 0.131 & 0.029  \\ 
    -0.0625 & -0.066 & 0.026 & 0.018 & 0.030  \\ 
    0.0625 & -0.005 & 0.026 & 0.012 & 0.029  \\ 
    0.1875 & 0.021 & 0.026 & -0.142 & 0.029  \\ 
    0.3125 & 0.124 & 0.026 & -0.149 & 0.029  \\ 
    0.4375 & 0.051 & 0.025 & -0.120 & 0.028  \\ 
    0.5625 & 0.183 & 0.025 & -0.090 & 0.027  \\ 
    0.6875 & 0.227 & 0.025 & -0.063 & 0.026  \\ 
    0.8125 & 0.220 & 0.026 & -0.041 & 0.026  \\ 
    0.9375 & 0.167 & 0.028 & 0.039 & 0.028  \\ 

    \hline
  \end{tabular}
\end{table*}

\begin{table*}
  \centering
  \caption{$\mathrm{\Lambda}$ polarization, dependence on Feynman x$_{\mathrm{F}}$ }
  \label{tablePol2}
  \begin{tabular}{c|cc|cc}
    &\multicolumn{2} {c|}{p$_{\mathrm{b}}=2.70 $\,GeV/c}  &\multicolumn{2} {c}{ p$_{\mathrm{b}}=2.95 $\,GeV/c }  \\
    x$_{\mathrm{F}}$                       & polarization                   & $\pm$&polarization                    &  $\pm$    \\
    \hline
    -0.9375 & -0.340 & 0.166 & 0.153 & 0.147  \\ 
    -0.8125 & -0.108 & 0.056 & -0.019 & 0.053  \\ 
    -0.6875 & -0.269 & 0.036 & -0.010 & 0.035  \\ 
    -0.5625 & -0.202 & 0.028 & 0.074 & 0.028  \\ 
    -0.4375 & -0.180 & 0.023 & 0.063 & 0.024  \\ 
    -0.3125 & -0.147 & 0.021 & 0.100 & 0.022  \\ 
    -0.1875 & -0.076 & 0.020 & 0.073 & 0.021  \\ 
    -0.0625 & -0.082 & 0.019 & 0.053 & 0.021  \\ 
    0.0625 & -0.000 & 0.019 & -0.029 & 0.021  \\ 
    0.1875 & 0.080 & 0.020 & -0.101 & 0.022  \\ 
    0.3125 & 0.110 & 0.021 & -0.131 & 0.023  \\ 
    0.4375 & 0.212 & 0.024 & -0.084 & 0.025  \\ 
    0.5625 & 0.245 & 0.027 & -0.044 & 0.029  \\ 
    0.6875 & 0.190 & 0.034 & -0.054 & 0.036  \\ 
    0.8125 & 0.260 & 0.051 & 0.111 & 0.050  \\ 
    0.9375 & 0.621 & 0.121 & 0.211 & 0.115  \\ 
    \hline
  \end{tabular}
\end{table*}

\begin{table*}
  \centering
  \caption{$\mathrm{\Lambda}$ polarization, dependence on the transversal momentum}
  \label{tablePol3}
  \begin{tabular}{ccc|ccc}
    \multicolumn{3} {c|}{p$_{\mathrm{b}}=2.70 $\,GeV/c}  &\multicolumn{3} {c}{ p$_{\mathrm{b}}=2.95 $\,GeV/c }  \\
    p$_{\mathrm{t}}$  [GeV/c]    & polarization   & $\pm$ &  p$_{\mathrm{t}}$  [GeV/c]  &polarization  &  $\pm$    \\
    \hline

    0.016 & -0.092 & 0.063 & 0.019 & 0.069 & 0.065  \\ 
    0.047 & 0.028 & 0.035 & 0.056 & 0.015 & 0.037  \\ 
    0.078 & 0.070 & 0.027 & 0.094 & 0.042 & 0.028  \\ 
    0.109 & 0.112 & 0.022 & 0.131 & -0.038 & 0.023  \\ 
    0.141 & 0.086 & 0.019 & 0.169 & -0.040 & 0.020  \\ 
    0.172 & 0.106 & 0.018 & 0.206 & -0.036 & 0.019  \\ 
    0.203 & 0.127 & 0.018 & 0.244 & -0.062 & 0.019  \\ 
    0.234 & 0.170 & 0.018 & 0.281 & -0.077 & 0.020  \\ 
    0.266 & 0.166 & 0.020 & 0.319 & -0.065 & 0.021  \\ 
    0.297 & 0.227 & 0.022 & 0.356 & -0.166 & 0.024  \\ 
    0.328 & 0.161 & 0.027 & 0.394 & -0.125 & 0.028  \\ 
    0.359 & 0.132 & 0.037 & 0.431 & -0.206 & 0.035  \\ 
    0.391 & 0.008 & 0.102 & 0.469 & -0.178 & 0.055  \\ 
    -    & -     & -     & 0.506 & -0.281 & 0.165  \\

    \hline
  \end{tabular}
\end{table*}

\clearpage

\begin{table*}
  \centering
  \caption{Analyzing power, 2.70\,GeV/c data dependence on the cm scattering angle  }
  \label{table3}
  \begin{tabular}{ccccccc}
    cos($\vartheta^{*}$)   & $A_{\mathrm{N}}(\varphi^{\mathrm{p}})$  &  $\pm$  & $A_{\mathrm{N}}(\varphi^{\mathrm{K}})$  &  $\pm$  & $A_{\mathrm{N}}(\varphi^{\mathrm{\Lambda}})$ &  $\pm$ \\

    \hline

    -0.9375 & -0.002 & 0.014 & 0.080 & 0.019 & -0.001 & 0.017  \\ 
    -0.8125 & -0.003 & 0.014 & 0.131 & 0.018 & -0.046 & 0.016  \\ 
    -0.6875 & -0.018 & 0.015 & 0.177 & 0.017 & -0.038 & 0.016  \\ 
    -0.5625 & -0.055 & 0.016 & 0.155 & 0.017 & -0.066 & 0.016  \\ 
    -0.4375 & -0.055 & 0.016 & 0.181 & 0.016 & -0.076 & 0.016  \\ 
    -0.3125 & -0.046 & 0.017 & 0.141 & 0.016 & -0.069 & 0.016  \\ 
    -0.1875 & -0.085 & 0.017 & 0.151 & 0.016 & -0.029 & 0.016  \\ 
    -0.0625 & -0.076 & 0.018 & 0.156 & 0.016 & -0.083 & 0.017  \\ 
    0.0625 & -0.029 & 0.018 & 0.129 & 0.016 & -0.045 & 0.016  \\ 
    0.1875 & -0.007 & 0.018 & 0.107 & 0.016 & -0.069 & 0.017  \\ 
    0.3125 & -0.019 & 0.018 & 0.090 & 0.016 & -0.047 & 0.017  \\ 
    0.4375 & -0.071 & 0.018 & 0.071 & 0.016 & -0.107 & 0.016  \\ 
    0.5625 & -0.016 & 0.017 & 0.046 & 0.016 & -0.107 & 0.016  \\ 
    0.6875 & -0.056 & 0.016 & 0.016 & 0.016 & -0.112 & 0.016  \\ 
    0.8125 & -0.011 & 0.015 & 0.070 & 0.016 & -0.065 & 0.016  \\ 
    0.9375 & -0.023 & 0.016 & 0.052 & 0.017 & -0.089 & 0.017  \\ 

    \hline
  \end{tabular}
\end{table*}

\begin{table*}
  \centering
  \caption{Analyzing power, 2.95\,GeV/c data dependence on the cm scattering angle  }
  \label{table4}
  \begin{tabular}{ccccccc}
    
    cos($\vartheta^{*}$)   & $A_{\mathrm{N}}(\varphi^{\mathrm{p}})$  &  $\pm$  & $A_{\mathrm{N}}(\varphi^{\mathrm{K}})$  &  $\pm$  & $A_{\mathrm{N}}(\varphi^{\mathrm{\Lambda}})$ &  $\pm$ \\
    \hline

    -0.9375 & 0.015 & 0.016 & 0.071 & 0.024 & -0.034 & 0.017  \\ 
    -0.8125 & 0.019 & 0.016 & 0.122 & 0.023 & -0.065 & 0.017  \\ 
    -0.6875 & 0.013 & 0.018 & 0.171 & 0.020 & -0.057 & 0.017  \\ 
    -0.5625 & -0.052 & 0.018 & 0.189 & 0.019 & -0.129 & 0.018  \\ 
    -0.4375 & -0.088 & 0.019 & 0.188 & 0.018 & -0.119 & 0.019  \\ 
    -0.3125 & -0.084 & 0.019 & 0.135 & 0.018 & -0.100 & 0.019  \\ 
    -0.1875 & -0.022 & 0.019 & 0.150 & 0.018 & -0.057 & 0.020  \\ 
    -0.0625 & -0.095 & 0.019 & 0.189 & 0.018 & -0.048 & 0.020  \\ 
    0.0625 & -0.078 & 0.020 & 0.149 & 0.018 & -0.020 & 0.019  \\ 
    0.1875 & -0.089 & 0.020 & 0.089 & 0.018 & -0.015 & 0.020  \\ 
    0.3125 & -0.075 & 0.020 & 0.060 & 0.018 & -0.011 & 0.019  \\ 
    0.4375 & -0.087 & 0.020 & 0.070 & 0.018 & 0.048 & 0.019  \\ 
    0.5625 & -0.002 & 0.020 & -0.003 & 0.018 & 0.001 & 0.019  \\ 
    0.6875 & 0.005 & 0.019 & -0.003 & 0.018 & 0.009 & 0.018  \\ 
    0.8125 & 0.041 & 0.018 & -0.035 & 0.018 & 0.066 & 0.017  \\ 
    0.9375 & 0.058 & 0.019 & -0.000 & 0.022 & 0.027 & 0.018  \\ 

    \hline
  \end{tabular}
\end{table*}
\clearpage

\begin{table*}
  \centering
  \caption{Analyzing power, 2.70\,GeV/c data dependence on Feynman  x$_{\mathrm{F}}$  }
  \label{table5}
  \begin{tabular}{ccccccc}
    x$_{\mathrm{F}}$    & $A_{\mathrm{N}}(\varphi^{\mathrm{p}})$  &  $\pm$  & $A_{\mathrm{N}}(\varphi^{\mathrm{K}})$  &  $\pm$  & $A_{\mathrm{N}}(\varphi^{\mathrm{\Lambda}})$ &  $\pm$ \\
    
    \hline

    -0.9375 & -0.014 & 0.056 & 0.144 & 0.081 & 0.110 & 0.096  \\ 
    -0.8125 & -0.002 & 0.024 & 0.118 & 0.033 & 0.043 & 0.034  \\ 
    -0.6875 & -0.018 & 0.017 & 0.207 & 0.022 & -0.000 & 0.022  \\ 
    -0.5625 & -0.051 & 0.015 & 0.181 & 0.017 & -0.056 & 0.017  \\ 
    -0.4375 & -0.016 & 0.014 & 0.133 & 0.015 & -0.056 & 0.015  \\ 
    -0.3125 & -0.026 & 0.013 & 0.144 & 0.014 & -0.069 & 0.013  \\ 
    -0.1875 & -0.062 & 0.013 & 0.143 & 0.013 & -0.047 & 0.012  \\ 
    -0.0625 & -0.057 & 0.013 & 0.136 & 0.013 & -0.065 & 0.012  \\ 
    0.0625 & -0.024 & 0.013 & 0.119 & 0.013 & -0.062 & 0.012  \\ 
    0.1875 & -0.030 & 0.014 & 0.110 & 0.013 & -0.057 & 0.013  \\ 
    0.3125 & -0.042 & 0.015 & 0.077 & 0.014 & -0.089 & 0.014  \\ 
    0.4375 & -0.051 & 0.015 & 0.042 & 0.014 & -0.085 & 0.015  \\ 
    0.5625 & -0.031 & 0.017 & 0.044 & 0.015 & -0.141 & 0.017  \\ 
    0.6875 & -0.011 & 0.020 & 0.031 & 0.018 & -0.067 & 0.021  \\ 
    0.8125 & 0.031 & 0.027 & 0.023 & 0.026 & -0.093 & 0.031  \\ 
    0.9375 & -0.005 & 0.068 & -0.041 & 0.071 & -0.090 & 0.073  \\ 
    \hline

  \end{tabular}
\end{table*}

\begin{table*}
  \centering
  \caption{Analyzing power, 2.95\,GeV/c data dependence on Feynman  x$_{\mathrm{F}}$}
    \label{table6}
    \begin{tabular}{ccccccc}
      x$_{\mathrm{F}}$    & $A_{\mathrm{N}}(\varphi^{\mathrm{p}})$  &  $\pm$  & $A_{\mathrm{N}}(\varphi^{\mathrm{K}})$  &  $\pm$  & $A_{\mathrm{N}}(\varphi^{\mathrm{\Lambda}})$ &  $\pm$ \\

      \hline
      -0.9375 & 0.030 & 0.064 & -0.091 & 0.123 & -0.311 & 0.093  \\ 
      -0.8125 & 0.007 & 0.026 & 0.185 & 0.050 & -0.105 & 0.035  \\ 
      -0.6875 & 0.021 & 0.019 & 0.134 & 0.028 & -0.043 & 0.023  \\ 
      -0.5625 & -0.009 & 0.017 & 0.155 & 0.020 & -0.108 & 0.019  \\ 
      -0.4375 & -0.044 & 0.016 & 0.164 & 0.017 & -0.112 & 0.016  \\ 
      -0.3125 & -0.054 & 0.015 & 0.175 & 0.015 & -0.089 & 0.015  \\ 
      -0.1875 & -0.043 & 0.015 & 0.130 & 0.014 & -0.066 & 0.014  \\ 
      -0.0625 & -0.051 & 0.015 & 0.173 & 0.014 & -0.029 & 0.014  \\ 
      0.0625 & -0.067 & 0.015 & 0.157 & 0.014 & -0.023 & 0.014  \\ 
      0.1875 & -0.105 & 0.016 & 0.073 & 0.015 & -0.000 & 0.015  \\ 
      0.3125 & -0.025 & 0.017 & 0.070 & 0.015 & 0.009 & 0.016  \\ 
      0.4375 & 0.004 & 0.018 & 0.007 & 0.016 & 0.036 & 0.017  \\ 
      0.5625 & 0.035 & 0.020 & -0.042 & 0.018 & 0.076 & 0.019  \\ 
      0.6875 & 0.019 & 0.022 & -0.064 & 0.021 & 0.076 & 0.024  \\ 
      0.8125 & 0.073 & 0.029 & -0.144 & 0.031 & -0.055 & 0.033  \\ 
      0.9375 & 0.122 & 0.063 & 0.032 & 0.079 & 0.080 & 0.075  \\ 
      \hline

    \end{tabular}
\end{table*}
\clearpage

\begin{table*}
  \centering
  \caption{Analyzing power, 2.70\,GeV/c data dependence on transversal momentum  }
  \label{table7}
  \begin{tabular}{ccccccc}
    p$_{\mathrm{t}}$  [GeV/c]  & $A_{\mathrm{N}}(\varphi^{\mathrm{p}})$  &  $\pm$  & $A_{\mathrm{N}}(\varphi^{\mathrm{K}})$  &  $\pm$  & $A_{\mathrm{N}}(\varphi^{\mathrm{\Lambda}})$ &  $\pm$ \\

    \hline
    0.016 & 0.071 & 0.108 & 0.143 & 0.048 & -0.064 & 0.038  \\ 
    0.047 & -0.008 & 0.022 & 0.030 & 0.020 & -0.069 & 0.022  \\ 
    0.078 & -0.016 & 0.014 & 0.067 & 0.015 & -0.029 & 0.017  \\ 
    0.109 & -0.028 & 0.012 & 0.087 & 0.013 & -0.032 & 0.014  \\ 
    0.141 & 0.002 & 0.011 & 0.113 & 0.012 & -0.064 & 0.012  \\ 
    0.172 & -0.035 & 0.011 & 0.116 & 0.011 & -0.073 & 0.011  \\ 
    0.203 & -0.035 & 0.011 & 0.110 & 0.011 & -0.066 & 0.011  \\ 
    0.234 & -0.038 & 0.012 & 0.142 & 0.011 & -0.088 & 0.011  \\ 
    0.266 & -0.049 & 0.013 & 0.114 & 0.011 & -0.086 & 0.012  \\ 
    0.297 & -0.056 & 0.016 & 0.117 & 0.013 & -0.057 & 0.014  \\ 
    0.328 & -0.086 & 0.020 & 0.122 & 0.042 & -0.080 & 0.017  \\ 
    0.359 & -0.113 & 0.027 & -     & -     & -0.039 & 0.023  \\ 
    0.391 & -0.115 & 0.117 & -     & -     & -0.045 & 0.064  \\ 
    \hline
  \end{tabular}
\end{table*}

\begin{table*}
  \centering
  \caption{Analyzing power, 2.95\,GeV/c data dependence on transversal momentum  }
  \label{table8}
  \begin{tabular}{ccccccc}
    p$_{\mathrm{t}}$ [GeV/c]   & $A_{\mathrm{N}}(\varphi^{\mathrm{p}})$  &  $\pm$  & $A_{\mathrm{N}}(\varphi^{\mathrm{K}})$  &  $\pm$  & $A_{\mathrm{N}}(\varphi^{\mathrm{\Lambda}})$ &  $\pm$ \\

    \hline
    0.019 & 0.111 & 0.113 & 0.094 & 0.061 & -0.014 & 0.041  \\ 
    0.056 & -0.057 & 0.032 & 0.019 & 0.027 & 0.033 & 0.024  \\ 
    0.094 & 0.023 & 0.020 & 0.066 & 0.020 & 0.017 & 0.018  \\ 
    0.131 & 0.002 & 0.016 & 0.112 & 0.017 & -0.009 & 0.015  \\ 
    0.169 & 0.026 & 0.014 & 0.117 & 0.015 & -0.015 & 0.014  \\ 
    0.206 & 0.002 & 0.014 & 0.129 & 0.013 & 0.001 & 0.013  \\ 
    0.244 & 0.020 & 0.014 & 0.114 & 0.012 & -0.018 & 0.013  \\ 
    0.281 & 0.002 & 0.014 & 0.091 & 0.012 & -0.048 & 0.013  \\ 
    0.319 & -0.039 & 0.014 & 0.085 & 0.012 & -0.081 & 0.014  \\ 
    0.356 & -0.066 & 0.015 & 0.080 & 0.015 & -0.059 & 0.016  \\ 
    0.394 & -0.111 & 0.016 & 0.067 & 0.019 & -0.073 & 0.019  \\ 
    0.431 & -0.169 & 0.019 & 0.131 & 0.048 & -0.089 & 0.024  \\ 
    0.469 & -0.128 & 0.026 & -     & -     & -0.103 & 0.037  \\ 
    0.506 & -0.148 & 0.084 & -     & -     & -0.221 & 0.110  \\ 
    \hline
  \end{tabular}
\end{table*}
\clearpage

\begin{table*}
  \centering
  \caption{$D_{\mathrm{NN}}$, dependence on $\mathrm{\Lambda}$ cm scattering angle }
  \label{tableDnn1}
  \begin{tabular}{c|cc|cc}
    &\multicolumn{2} {c|}{p$_{\mathrm{b}}=2.70 $\,GeV/c}  &\multicolumn{2} {c}{ p$_{\mathrm{b}}=2.95 $\,GeV/c }  \\
    cos${\mathrm{\vartheta}_{\mathrm{\Lambda}}^{*}}$   & $D_{\mathrm{NN}}$ & $\pm$  & $D_{\mathrm{NN}}$ & $\pm$    \\
    \hline
    -0.9375 & -0.645 & 0.070 & -0.172 & 0.070  \\ 
    -0.8125 & -0.384 & 0.067 & -0.179 & 0.070  \\ 
    -0.6875 & -0.523 & 0.066 & -0.263 & 0.072  \\ 
    -0.5625 & -0.473 & 0.065 & -0.453 & 0.074  \\ 
    -0.4375 & -0.347 & 0.065 & -0.435 & 0.076  \\ 
    -0.3125 & -0.445 & 0.065 & -0.485 & 0.077  \\ 
    -0.1875 & -0.365 & 0.066 & -0.275 & 0.080  \\ 
    -0.0625 & -0.226 & 0.067 & -0.260 & 0.080  \\ 
    0.0625 & -0.305 & 0.066 & -0.318 & 0.079  \\ 
    0.1875 & -0.321 & 0.066 & -0.182 & 0.080  \\ 
    0.3125 & -0.215 & 0.066 & -0.383 & 0.078  \\ 
    0.4375 & -0.315 & 0.064 & -0.323 & 0.076  \\ 
    0.5625 & -0.430 & 0.064 & -0.252 & 0.075  \\ 
    0.6875 & -0.349 & 0.065 & -0.420 & 0.071  \\ 
    0.8125 & -0.455 & 0.065 & -0.407 & 0.070  \\ 
    0.9375 & -0.498 & 0.072 & -0.256 & 0.075  \\ 
    \hline

  \end{tabular}
\end{table*}

\begin{table*}
  \centering
  \caption{ $D_{\mathrm{NN}}$, dependence on Feynman x$_{\mathrm{F}}$ }
  \label{tableDnn2}
  \begin{tabular}{c|cc|cc}
    &\multicolumn{2} {c|}{p$_{\mathrm{b}}=2.70 $\,GeV/c}  &\multicolumn{2} {c}{ p$_{\mathrm{b}}=2.95 $\,GeV/c }  \\
    x$_{\mathrm{F}}$                       &   $D_{\mathrm{NN}}$                 & $\pm$& $D_{\mathrm{NN}}$                    &  $\pm$    \\
    \hline

    -0.9375 & 0.181 & 0.422 & 0.206 & 0.401  \\ 
    -0.8125 & -0.636 & 0.140 & 0.042 & 0.144  \\ 
    -0.6875 & -0.306 & 0.091 & 0.040 & 0.096  \\ 
    -0.5625 & -0.407 & 0.070 & -0.268 & 0.076  \\ 
    -0.4375 & -0.497 & 0.059 & -0.284 & 0.066  \\ 
    -0.3125 & -0.447 & 0.054 & -0.313 & 0.061  \\ 
    -0.1875 & -0.469 & 0.050 & -0.559 & 0.058  \\ 
    -0.0625 & -0.333 & 0.049 & -0.305 & 0.057  \\ 
    0.0625 & -0.349 & 0.049 & -0.259 & 0.057  \\ 
    0.1875 & -0.235 & 0.050 & -0.323 & 0.059  \\ 
    0.3125 & -0.306 & 0.054 & -0.243 & 0.062  \\ 
    0.4375 & -0.517 & 0.060 & -0.391 & 0.068  \\ 
    0.5625 & -0.379 & 0.069 & -0.410 & 0.078  \\ 
    0.6875 & -0.530 & 0.086 & -0.355 & 0.096  \\ 
    0.8125 & -0.348 & 0.128 & -0.426 & 0.134  \\ 
    0.9375 & -0.285 & 0.312 & -0.452 & 0.313  \\ 

    \hline
  \end{tabular}
\end{table*}

\begin{table*}
  \centering
  \caption{$D_{\mathrm{NN}}$, dependence on the transversal momentum}
  \label{tableDnn3}
  \begin{tabular}{ccc|ccc}
    \multicolumn{3} {c|}{p$_{\mathrm{b}}=2.70 $\,GeV/c}  &\multicolumn{3} {c}{ p$_{\mathrm{b}}=2.95 $\,GeV/c }  \\
    p$_{\mathrm{t}}$  [GeV/c]    &$D_{\mathrm{NN}}$    & $\pm$ &  p$_{\mathrm{t}}$  [GeV/c]  &$D_{\mathrm{NN}}$   &  $\pm$    \\
    \hline
    0.016 & -0.784 & 0.157 & 0.019 & -0.231 & 0.172  \\ 
    0.047 & -0.707 & 0.089 & 0.056 & -0.217 & 0.098  \\ 
    0.078 & -0.505 & 0.068 & 0.094 & -0.408 & 0.075  \\ 
    0.109 & -0.503 & 0.056 & 0.131 & -0.338 & 0.061  \\ 
    0.141 & -0.438 & 0.048 & 0.169 & -0.298 & 0.055  \\ 
    0.172 & -0.460 & 0.045 & 0.206 & -0.350 & 0.053  \\ 
    0.203 & -0.319 & 0.045 & 0.244 & -0.337 & 0.052  \\ 
    0.234 & -0.370 & 0.046 & 0.281 & -0.292 & 0.054  \\ 
    0.266 & -0.457 & 0.050 & 0.319 & -0.366 & 0.058  \\ 
    0.297 & -0.266 & 0.056 & 0.356 & -0.294 & 0.064  \\ 
    0.328 & -0.034 & 0.067 & 0.394 & -0.246 & 0.076  \\ 
    0.359 & -0.160 & 0.094 & 0.431 & -0.253 & 0.096  \\ 
    0.391 & 0.034 & 0.258 & 0.469 & -0.244 & 0.150  \\ 
    -     &     - &     - & 0.506 & 0.000 & 0.456  \\ 

    \hline
  \end{tabular}
\end{table*}

\clearpage

\clearpage
\bibliography{bibForPol.bbl}
\clearpage
\end{document}